
%
%
%
\def\unredoffs{} \def\redoffs{\voffset=-.31truein\hoffset=-.48truein}
\def\speclscape{}
%
%
%
%
%
\newbox\leftpage \newdimen\fullhsize \newdimen\hstitle \newdimen\hsbody
\tolerance=1000\hfuzz=2pt
\catcode`\@=11 
\ifx\hyperdef\UNd@FiNeD\def\hyperdef#1#2#3#4{#4}\def\hyperref#1#2#3#4{#4}\fi
\def\bigans{b }
\def\answ{b }
%
\ifx\answ\bigans\message{(This will come out unreduced.}
\magnification=1200\unredoffs\baselineskip=16pt plus 2pt minus 1pt
\hsbody=\hsize \hstitle=\hsize 
\else\message{(This will be reduced.} \let\l@r=L
\magnification=1000\baselineskip=16pt plus 2pt minus 1pt \vsize=7truein
\redoffs \hstitle=8truein\hsbody=4.75truein\fullhsize=10truein\hsize=\hsbody
\output={\ifnum\pageno=0 
  \shipout\vbox{\speclscape{\hsize\fullhsize\makeheadline}
    \hbox to \fullhsize{\hfill\pagebody\hfill}}\advancepageno
  \else
  \almostshipout{\leftline{\vbox{\pagebody\makefootline}}}\advancepageno
  \fi}
\def\almostshipout#1{\if L\l@r \count1=1 \message{[\the\count0.\the\count1]}
      \global\setbox\leftpage=#1 \global\let\l@r=R
 \else \count1=2
  \shipout\vbox{\speclscape{\hsize\fullhsize\makeheadline}
      \hbox to\fullhsize{\box\leftpage\hfil#1}}  \global\let\l@r=L\fi}
\fi
%
\newcount\yearltd\yearltd=\year\advance\yearltd by -2000

\def\Title#1#2{\nopagenumbers\abstractfont\hsize=\hstitle\rightline{#1}%
\vskip 1in\centerline{\titlefont #2}\abstractfont\vskip .5in\pageno=0}
\def\Date#1{\vfill\leftline{#1}\tenpoint\supereject\global\hsize=\hsbody%
\footline={\hss\tenrm\hyperdef\hypernoname{page}\folio\folio\hss}}%
%

\def\draftmode{\message{ DRAFTMODE }\def\draftdate{{\rm preliminary draft:
\number\month/\number\day/\number\yearltd\ \ \hourmin}}%
\headline={\hfil\draftdate}\writelabels\baselineskip=20pt plus 2pt minus 2pt
 {\count255=\time\divide\count255 by 60 \xdef\hourmin{\number\count255}
  \multiply\count255 by-60\advance\count255 by\time
  \xdef\hourmin{\hourmin:\ifnum\count255<10 0\fi\the\count255}}}
\def\nolabels{\def\wrlabeL##1{}\def\eqlabeL##1{}\def\reflabeL##1{}}
\def\writelabels{\def\wrlabeL##1{\leavevmode\vadjust{\rlap{\smash%
{\line{{\escapechar=` \hfill\rlap{\sevenrm\hskip.03in\string##1}}}}}}}%
\def\eqlabeL##1{{\escapechar-1\rlap{\sevenrm\hskip.05in\string##1}}}%
\def\reflabeL##1{\noexpand\llap{\noexpand\sevenrm\string\string\string##1}}}
\nolabels
%
\global\newcount\secno \global\secno=0
\global\newcount\meqno \global\meqno=1
\def\s@csym{}
\def\newsec#1{\global\advance\secno by1%
{\toks0{#1}\message{(\the\secno. \the\toks0)}}%
\global\subsecno=0\eqnres@t\let\s@csym\secsym\xdef\secn@m{\the\secno}\noindent
{\bf\hyperdef\hypernoname{section}{\the\secno}{\the\secno.} #1}%
\writetoca{{\string\hyperref{}{section}{\the\secno}{\the\secno.}} {#1}}%
\par\nobreak\medskip\nobreak}
\def\eqnres@t{\xdef\secsym{\the\secno.}\global\meqno=1\bigbreak\bigskip}
\def\sequentialequations{\def\eqnres@t{\bigbreak}}\xdef\secsym{}
\global\newcount\subsecno \global\subsecno=0
\def\subsec#1{\global\advance\subsecno by1%
{\toks0{#1}\message{(\s@csym\the\subsecno. \the\toks0)}}%
\ifnum\lastpenalty>9000\else\bigbreak\fi
\noindent{\it\hyperdef\hypernoname{subsection}{\secn@m.\the\subsecno}%
{\secn@m.\the\subsecno.} #1}\writetoca{\string\quad
{\string\hyperref{}{subsection}{\secn@m.\the\subsecno}{\secn@m.\the\subsecno.}}
{#1}}\par\nobreak\medskip\nobreak}
\def\appendix#1#2{\global\meqno=1\global\subsecno=0\xdef\secsym{\hbox{#1.}}%
\bigbreak\bigskip\noindent{\bf Appendix \hyperdef\hypernoname{appendix}{#1}%
{#1.} #2}{\toks0{(#1. #2)}\message{\the\toks0}}%
\xdef\s@csym{#1.}\xdef\secn@m{#1}%
\writetoca{\string\hyperref{}{appendix}{#1}{Appendix {#1.}} {#2}}%
\par\nobreak\medskip\nobreak}
%
%
\def\checkm@de#1#2{\ifmmode{\def\f@rst##1{##1}\hyperdef\hypernoname{equation}%
{#1}{#2}}\else\hyperref{}{equation}{#1}{#2}\fi}
\def\eqnn#1{\DefWarn#1\xdef #1{(\noexpand\relax\noexpand\checkm@de%
{\s@csym\the\meqno}{\secsym\the\meqno})}%
\wrlabeL#1\writedef{#1\leftbracket#1}\global\advance\meqno by1}
\def\f@rst#1{\c@t#1a\em@ark}\def\c@t#1#2\em@ark{#1}
\def\eqna#1{\DefWarn#1\wrlabeL{#1$\{\}$}%
\xdef #1##1{(\noexpand\relax\noexpand\checkm@de%
{\s@csym\the\meqno\noexpand\f@rst{##1}}{\hbox{$\secsym\the\meqno##1$}})}
\writedef{#1\numbersign1\leftbracket#1{\numbersign1}}\global\advance\meqno by1}
\def\eqn#1#2{\DefWarn#1%
\xdef #1{(\noexpand\hyperref{}{equation}{\s@csym\the\meqno}%
{\secsym\the\meqno})}$$#2\eqno(\hyperdef\hypernoname{equation}%
{\s@csym\the\meqno}{\secsym\the\meqno})\eqlabeL#1$$%
\writedef{#1\leftbracket#1}\global\advance\meqno by1}
\def\xeqn{\expandafter\xe@n}\def\xe@n(#1){#1}
\def\xeqna#1{\expandafter\xe@n#1}
\def\eqns#1{(\e@ns #1{\hbox{}})}
\def\e@ns#1{\ifx\UNd@FiNeD#1\message{eqnlabel \string#1 is undefined.}%
\xdef#1{(?.?)}\fi{\let\hyperref=\relax\xdef\next{#1}}%
\ifx\next\em@rk\def\next{}\else%
\ifx\next#1\xeqn#1\else\def\n@xt{#1}\ifx\n@xt\next#1\else\xeqna#1\fi
\fi\let\next=\e@ns\fi\next}

\def\DefWarn#1{\ifx\UNd@FiNeD#1\else
\immediate\write16{*** WARNING: the label \string#1 is already defined ***}\fi}
%
\newskip\footskip\footskip14pt plus 1pt minus 1pt 
\def\footnotefont{\ninepoint}\def\f@t#1{\footnotefont #1\@foot}
\def\f@@t{\baselineskip\footskip\bgroup\footnotefont\aftergroup\@foot\let\next}
\setbox\strutbox=\hbox{\vrule height9.5pt depth4.5pt width0pt}
\global\newcount\ftno \global\ftno=0
\def\foot{\global\advance\ftno by1\def\foot@rg{\hyperref{}{footnote}%
{\the\ftno}{\the\ftno}\xdef\foot@rg{\noexpand\hyperdef\noexpand\hypernoname%
{footnote}{\the\ftno}{\the\ftno}}}\footnote{$^{\foot@rg}$}}
%
\newwrite\ftfile
\def\footend{\def\foot{\global\advance\ftno by1\chardef\wfile=\ftfile
\hyperref{}{footnote}{\the\ftno}{$^{\the\ftno}$}%
\ifnum\ftno=1\immediate\openout\ftfile=\jobname.fts\fi%
\immediate\write\ftfile{\noexpand\smallskip%
\noexpand\item{\noexpand\hyperdef\noexpand\hypernoname{footnote}
{\the\ftno}{f\the\ftno}:\ }\pctsign}\findarg}%
\def\footatend{\vfill\eject\immediate\closeout\ftfile{\parindent=20pt
\centerline{\bf Footnotes}\nobreak\bigskip\input \jobname.fts }}}
\def\footatend{}
%
%
\global\newcount\refno \global\refno=1
\newwrite\rfile
\def\ref{[\hyperref{}{reference}{\the\refno}{\the\refno}]\nref}
\def\nref#1{\DefWarn#1%
\xdef#1{[\noexpand\hyperref{}{reference}{\the\refno}{\the\refno}]}%
\writedef{#1\leftbracket#1}%
\ifnum\refno=1\immediate\openout\rfile=\jobname.refs\fi
\chardef\wfile=\rfile\immediate\write\rfile{\noexpand\item{[\noexpand\hyperdef%
\noexpand\hypernoname{reference}{\the\refno}{\the\refno}]\ }%
\reflabeL{#1\hskip.31in}\pctsign}\global\advance\refno by1\findarg}
\def\findarg#1#{\begingroup\obeylines\newlinechar=`\^^M\pass@rg}
{\obeylines\gdef\pass@rg#1{\writ@line\relax #1^^M\hbox{}^^M}%
\gdef\writ@line#1^^M{\expandafter\toks0\expandafter{\striprel@x #1}%
\edef\next{\the\toks0}\ifx\next\em@rk\let\next=\endgroup\else\ifx\next\empty%
\else\immediate\write\wfile{\the\toks0}\fi\let\next=\writ@line\fi\next\relax}}
\def\striprel@x#1{} \def\em@rk{\hbox{}}
\def\lref{\begingroup\obeylines\lr@f}
\def\lr@f#1#2{\DefWarn#1\gdef#1{\let#1=\UNd@FiNeD\ref#1{#2}}\endgroup\unskip}

\def\addref#1{\immediate\write\rfile{\noexpand\item{}#1}} 
\def\listrefs{\footatend\vfill\supereject\immediate\closeout\rfile\writestoppt
\baselineskip=\footskip\centerline{{\bf References}}\bigskip{\parindent=20pt%
\frenchspacing\escapechar=` \input \jobname.refs\vfill\eject}\nonfrenchspacing}
\def\startrefs#1{\immediate\openout\rfile=\jobname.refs\refno=#1}
\def\xref{\expandafter\xr@f}\def\xr@f[#1]{#1}
\def\refs#1{\count255=1[\r@fs #1{\hbox{}}]}
\def\r@fs#1{\ifx\UNd@FiNeD#1\message{reflabel \string#1 is undefined.}%
\nref#1{need to supply reference \string#1.}\fi%
\vphantom{\hphantom{#1}}{\let\hyperref=\relax\xdef\next{#1}}%
\ifx\next\em@rk\def\next{}%
\else\ifx\next#1\ifodd\count255\relax\xref#1\count255=0\fi%
\else#1\count255=1\fi\let\next=\r@fs\fi\next}
%

%
\newwrite\ffile\global\newcount\figno \global\figno=1
\def\fig{fig.~\hyperref{}{figure}{\the\figno}{\the\figno}\nfig}
\def\nfig#1{\DefWarn#1%
\xdef#1{fig.~\noexpand\hyperref{}{figure}{\the\figno}{\the\figno}}%
\writedef{#1\leftbracket fig.\noexpand~\xfig#1}%
\ifnum\figno=1\immediate\openout\ffile=\jobname.figs\fi\chardef\wfile=\ffile%
{\let\hyperref=\relax
\immediate\write\ffile{\noexpand\medskip\noexpand\item{Fig.\ %
\noexpand\hyperdef\noexpand\hypernoname{figure}{\the\figno}{\the\figno}. }
\reflabeL{#1\hskip.55in}\pctsign}}\global\advance\figno by1\findarg}
\def\listfigs{\vfill\eject\immediate\closeout\ffile{\parindent40pt
\baselineskip14pt\centerline{{\bf Figure Captions}}\nobreak\medskip
\escapechar=` \input \jobname.figs\vfill\eject}}
\def\xfig{\expandafter\xf@g}\def\xf@g fig.\penalty\@M\ {}
\def\figs#1{figs.~\f@gs #1{\hbox{}}}
\def\f@gs#1{{\let\hyperref=\relax\xdef\next{#1}}\ifx\next\em@rk\def\next{}\else
\ifx\next#1\xfig #1\else#1\fi\let\next=\f@gs\fi\next}
\def\figin{\epsfcheck\figin}\def\figins{\epsfcheck\figins}
\def\epsfcheck{\ifx\epsfbox\UNd@FiNeD
\message{(NO epsf.tex, FIGURES WILL BE IGNORED)}
\gdef\figin##1{\vskip2in}\gdef\figins##1{\hskip.5in}
\else\message{(FIGURES WILL BE INCLUDED)}%
\gdef\figin##1{##1}\gdef\figins##1{##1}\fi}
\def\DefWarn#1{}
\def\figinsert{\goodbreak\midinsert}
\def\ifig#1#2#3{\DefWarn#1\xdef#1{fig.~\noexpand\hyperref{}{figure}%
{\the\figno}{\the\figno}}\writedef{#1\leftbracket fig.\noexpand~\xfig#1}%
\figinsert\figin{\centerline{#3}}\medskip\centerline{\vbox{\baselineskip12pt
\advance\hsize by -1truein\noindent\wrlabeL{#1=#1}\footnotefont%
{\bf Fig.~\hyperdef\hypernoname{figure}{\the\figno}{\the\figno}:} #2}}
\bigskip\endinsert\global\advance\figno by1}
\newwrite\lfile
{\escapechar-1\xdef\pctsign{\string\%}\xdef\leftbracket{\string\{}
\xdef\rightbracket{\string\}}\xdef\numbersign{\string\#}}
\def\writedefs{\immediate\openout\lfile=\jobname.defs \def\writedef##1{%
{\let\hyperref=\relax\let\hyperdef=\relax\let\hypernoname=\relax
 \immediate\write\lfile{\string\def\string##1\rightbracket}}}}%
\def\writestop{\def\writestoppt{\immediate\write\lfile{\string\pageno
 \the\pageno\string\startrefs\leftbracket\the\refno\rightbracket
 \string\def\string\secsym\leftbracket\secsym\rightbracket
 \string\secno\the\secno\string\meqno\the\meqno}\immediate\closeout\lfile}}
\def\writestoppt{}\def\writedef#1{}
\def\seclab#1{\DefWarn#1%
\xdef #1{\noexpand\hyperref{}{section}{\the\secno}{\the\secno}}%
\writedef{#1\leftbracket#1}\wrlabeL{#1=#1}}
\def\subseclab#1{\DefWarn#1%
\xdef #1{\noexpand\hyperref{}{subsection}{\secn@m.\the\subsecno}%
{\secn@m.\the\subsecno}}\writedef{#1\leftbracket#1}\wrlabeL{#1=#1}}
\def\applab#1{\DefWarn#1%
\xdef #1{\noexpand\hyperref{}{appendix}{\secn@m}{\secn@m}}%
\writedef{#1\leftbracket#1}\wrlabeL{#1=#1}}
\newwrite\tfile \def\writetoca#1{}
\def\leaderfill{\leaders\hbox to 1em{\hss.\hss}\hfill}
\def\writetoc{\immediate\openout\tfile=\jobname.toc
   \def\writetoca##1{{\edef\next{\write\tfile{\noindent ##1
   \string\leaderfill {\string\hyperref{}{page}{\noexpand\number\pageno}%
                       {\noexpand\number\pageno}} \par}}\next}}}
\newread\ch@ckfile
\def\listtoc{\immediate\closeout\tfile\immediate\openin\ch@ckfile=\jobname.toc
\ifeof\ch@ckfile\message{no file \jobname.toc, no table of contents this pass}%
\else\closein\ch@ckfile\centerline{\bf Contents}\nobreak\medskip%
{\baselineskip=12pt\footnotefont\parskip=0pt\catcode`\@=11\input\jobname.toc
\catcode`\@=12\bigbreak\bigskip}\fi}
\catcode`\@=12 
%
\edef\tfontsize{\ifx\answ\bigans scaled\magstep3\else scaled\magstep4\fi}
\font\titlerm=cmr10 \tfontsize \font\titlerms=cmr7 \tfontsize
\font\titlermss=cmr5 \tfontsize \font\titlei=cmmi10 \tfontsize
\font\titleis=cmmi7 \tfontsize \font\titleiss=cmmi5 \tfontsize
\font\titlesy=cmsy10 \tfontsize \font\titlesys=cmsy7 \tfontsize
\font\titlesyss=cmsy5 \tfontsize \font\titleit=cmti10 \tfontsize
\skewchar\titlei='177 \skewchar\titleis='177 \skewchar\titleiss='177
\skewchar\titlesy='60 \skewchar\titlesys='60 \skewchar\titlesyss='60
\def\titlefont{\def\rm{\fam0\titlerm}
\textfont0=\titlerm \scriptfont0=\titlerms \scriptscriptfont0=\titlermss
\textfont1=\titlei \scriptfont1=\titleis \scriptscriptfont1=\titleiss
\textfont2=\titlesy \scriptfont2=\titlesys \scriptscriptfont2=\titlesyss
\textfont\itfam=\titleit \def\it{\fam\itfam\titleit}\rm}
 \ifx\answ\bigans\else scaled\magstep1\fi
\ifx\answ\bigans\def\abstractfont{\tenpoint}\else
\font\absit=cmti10 scaled \magstep1
\font\abssl=cmsl10 scaled \magstep1
\font\absrm=cmr10 scaled\magstep1 \font\absrms=cmr7 scaled\magstep1
\font\absrmss=cmr5 scaled\magstep1 \font\absi=cmmi10 scaled\magstep1
\font\absis=cmmi7 scaled\magstep1 \font\absiss=cmmi5 scaled\magstep1
\font\abssy=cmsy10 scaled\magstep1 \font\abssys=cmsy7 scaled\magstep1
\font\abssyss=cmsy5 scaled\magstep1 \font\absbf=cmbx10 scaled\magstep1
\skewchar\absi='177 \skewchar\absis='177 \skewchar\absiss='177
\skewchar\abssy='60 \skewchar\abssys='60 \skewchar\abssyss='60
\def\abstractfont{\def\rm{\fam0\absrm}
\textfont0=\absrm \scriptfont0=\absrms \scriptscriptfont0=\absrmss
\textfont1=\absi \scriptfont1=\absis \scriptscriptfont1=\absiss
\textfont2=\abssy \scriptfont2=\abssys \scriptscriptfont2=\abssyss
\textfont\itfam=\absit \def\it{\fam\itfam\absit}\def\footnotefont{\tenpoint}%
\textfont\slfam=\abssl \def\sl{\fam\slfam\abssl}%
\textfont\bffam=\absbf \def\bf
{\fam\bffam\absbf}\rm}\fi
\def\tenpoint{\def\rm{\fam0\tenrm}
\textfont0=\tenrm \scriptfont0=\sevenrm \scriptscriptfont0=\fiverm
\textfont1=\teni  \scriptfont1=\seveni  \scriptscriptfont1=\fivei
\textfont2=\tensy \scriptfont2=\sevensy \scriptscriptfont2=\fivesy
\textfont\itfam=\tenit \def\it{\fam\itfam\tenit}\def\footnotefont{\ninepoint}%
\textfont\bffam=\tenbf \def\bf{\fam\bffam\tenbf}\def\sl{\fam\slfam\tensl}\rm}
\font\ninerm=cmr9 \font\sixrm=cmr6 \font\ninei=cmmi9 \font\sixi=cmmi6
\font\ninesy=cmsy9 \font\sixsy=cmsy6 \font\ninebf=cmbx9
\font\nineit=cmti9 \font\ninesl=cmsl9 \skewchar\ninei='177
\skewchar\sixi='177 \skewchar\ninesy='60 \skewchar\sixsy='60
\def\ninepoint{\def\rm{\fam0\ninerm}
\textfont0=\ninerm \scriptfont0=\sixrm \scriptscriptfont0=\fiverm
\textfont1=\ninei \scriptfont1=\sixi \scriptscriptfont1=\fivei
\textfont2=\ninesy \scriptfont2=\sixsy \scriptscriptfont2=\fivesy
\textfont\itfam=\ninei \def\it{\fam\itfam\nineit}\def\sl{\fam\slfam\ninesl}%
\textfont\bffam=\ninebf \def\bf{\fam\bffam\ninebf}\rm}
%
%

\hyphenation{anom-aly anom-alies coun-ter-term coun-ter-terms}
\def\inv{^{\raise.15ex\hbox{${\scriptscriptstyle -}$}\kern-.05em 1}}

\def\Dsl{\,\raise.15ex\hbox{/}\mkern-13.5mu D} 
\def\dsl{\raise.15ex\hbox{/}\kern-.57em\partial}

 \def\Tr{{\rm Tr}}
\def\lspace{\ifx\answ\bigans{}\else\qquad\fi}
\def\lbspace{\ifx\answ\bigans{}\else\hskip-.2in\fi} 

\def\boxeqn#1{\vcenter{\vbox{\hrule\hbox{\vrule\kern3pt\vbox{\kern3pt
	\hbox{${\displaystyle #1}$}\kern3pt}\kern3pt\vrule}\hrule}}}
\def\mbox#1#2{\vcenter{\hrule \hbox{\vrule height#2in
		\kern#1in \vrule} \hrule}}  
%
 \def\CC{{\cal C}}

\def\vev#1{\langle #1 \rangle}

\def\darr#1{\raise1.5ex\hbox{$\leftrightarrow$}\mkern-16.5mu #1}

\def\roughly#1{\raise.3ex\hbox{$#1$\kern-.75em\lower1ex\hbox{$\sim$}}}

\def\bb{
\font\tenmsb=msbm10
\font\sevenmsb=msbm7
\font\fivemsb=msbm5
\textfont1=\tenmsb
\scriptfont1=\sevenmsb
\scriptscriptfont1=\fivemsb
}

\input amssym

\input epsf

\def\IZ{\relax\ifmmode\mathchoice
{\hbox{\cmss Z\kern-.4em Z}}{\hbox{\cmss Z\kern-.4em Z}} {\lower.9pt\hbox{\cmsss Z\kern-.4em Z}}
{\lower1.2pt\hbox{\cmsss Z\kern-.4em Z}}\else{\cmss Z\kern-.4em Z}\fi}

\newif\ifdraft\draftfalse
\newif\ifinter\interfalse
\ifdraft\draftmode\else\interfalse\fi
\def\journal#1&#2(#3){\unskip, \sl #1\ \bf #2 \rm(19#3) }
\def\andjournal#1&#2(#3){\sl #1~\bf #2 \rm (19#3) }

\def\frac#1#2{{#1\over#2}}

\def\vev#1{\langle#1\rangle}

\def\inbar{\,\vrule height1.5ex width.4pt depth0pt}
\def\IC{\relax\hbox{$\inbar\kern-.3em{\rm C}$}}
\def\IR{\relax{\rm I\kern-.18em R}}
\def\IP{\relax{\rm I\kern-.18em P}}

%
%


%
\catcode`\@=11
\def\slash#1{\mathord{\mathpalette\c@ncel{#1}}}
\overfullrule=0pt

\def\CC{{\cal C}}

\def\II{{\cal I}}

\def\MM{{\cal M}}
\def\NN{{\cal N}}

\def\PP{{\cal P}}

\def\S{\hbox{$\bb S$}}

\def\Z{\hbox{$\bb Z$}}
\def\R{\hbox{$\bb R$}}

\def\underrel#1\over#2{\mathrel{\mathop{\kern\z@#1}\limits_{#2}}}

\catcode`\@=12


%

\def\vev#1{\left\langle #1 \right\rangle}
\def\det{{\rm det}}

\def\det{{\rm det}}
\def\exp{{\rm exp}}


\def\[{[}
\def\]{]}

\def\comment#1{ }

%
\def\draftnote#1{\ifdraft{\baselineskip2ex
                 \vbox{\kern1em\hrule\hbox{\vrule\kern1em\vbox{\kern1ex
                 \noindent \underbar{NOTE}: #1
             \vskip1ex}\kern1em\vrule}\hrule}}\fi}
\def\internote#1{\ifinter{\baselineskip2ex
                 \vbox{\kern1em\hrule\hbox{\vrule\kern1em\vbox{\kern1ex
                 \noindent \underbar{Internal Note}: #1
             \vskip1ex}\kern1em\vrule}\hrule}}\fi}

%
%



%
%
%
%

%

\def\inv{^{-1}}


\def\Tr{{\rm Tr}}

\lref\DaviesNW{
  N.~M.~Davies, T.~J.~Hollowood and V.~V.~Khoze,
  ``Monopoles, affine algebras and the gluino condensate,''
  J.\ Math.\ Phys.\  {\bf 44}, 3640 (2003).'
  [hep-th/0006011].
}

\lref\KutasovVE{
  D.~Kutasov,
  ``A Comment on duality in N=1 supersymmetric nonAbelian gauge theories,''
Phys.\ Lett.\ B {\bf 351}, 230 (1995).
[hep-th/9503086].
}

\lref\KinneyEJ{
  J.~Kinney, J.~M.~Maldacena, S.~Minwalla and S.~Raju,
  ``An Index for 4 dimensional super conformal theories,''
Commun.\ Math.\ Phys.\  {\bf 275}, 209 (2007).
[hep-th/0510251].
}

\lref\DaviesUW{
  N.~M.~Davies, T.~J.~Hollowood, V.~V.~Khoze and M.~P.~Mattis,
  ``Gluino condensate and magnetic monopoles in supersymmetric gluodynamics,''
  Nucl.\ Phys.\ B {\bf 559}, 123 (1999).[hep-th/9905015].
  }

\lref\LeeVP{
  K.~-M.~Lee and P.~Yi,
  ``Monopoles and instantons on partially compactified D-branes,''
  Phys.\ Rev.\ D {\bf 56}, 3711 (1997).
  [hep-th/9702107].
  }

\lref\LeeVU{
  K.~-M.~Lee,
  ``Instantons and magnetic monopoles on R**3 x S**1 with arbitrary simple gauge groups,''
  Phys.\ Lett.\ B {\bf 426}, 323 (1998).[hep-th/9802012].
  }

\lref\NiarchosAA{
  V.~Niarchos,
  ``R-charges, Chiral Rings and RG Flows in Supersymmetric Chern-Simons-Matter Theories,''
JHEP {\bf 0905}, 054 (2009).
[arXiv:0903.0435 [hep-th]].
}

\lref\IntriligatorER{
  K.~A.~Intriligator and N.~Seiberg,
  ``Phases of N=1 supersymmetric gauge theories and electric - magnetic triality,''
In *Los Angeles 1995, Future perspectives in string theory* 270-282.
[hep-th/9506084].
}
\lref\NiarchosJB{
  V.~Niarchos,
  ``Seiberg Duality in Chern-Simons Theories with Fundamental and Adjoint Matter,''
JHEP {\bf 0811}, 001 (2008).
[arXiv:0808.2771 [hep-th]].
}

\lref\BorokhovIB{
  V.~Borokhov, A.~Kapustin and X.~-k.~Wu,
  ``Topological disorder operators in three-dimensional conformal field theory,''
JHEP {\bf 0211}, 049 (2002).
[hep-th/0206054].
}

\lref\ParkWTA{
  J.~Park and K.~-J.~Park,
  ``Seiberg-like Dualities for 3d N=2 Theories with SU(N) gauge group,''
[arXiv:1305.6280 [hep-th]].
}

\lref\BorokhovCG{
  V.~Borokhov, A.~Kapustin and X.~-k.~Wu,
  ``Monopole operators and mirror symmetry in three-dimensions,''
JHEP {\bf 0212}, 044 (2002).
[hep-th/0207074].
}

\lref\NakanishiHJ{
  T.~Nakanishi and A.~Tsuchiya,
  ``Level rank duality of WZW models in conformal field theory,''
Commun.\ Math.\ Phys.\  {\bf 144}, 351 (1992).
}
\lref\NiarchosAH{
  V.~Niarchos,
  ``Seiberg dualities and the 3d/4d connection,''
JHEP {\bf 1207}, 075 (2012).
[arXiv:1205.2086 [hep-th]].
}

\lref\SeibergBZ{
  N.~Seiberg,
  ``Exact results on the space of vacua of four-dimensional SUSY gauge theories,''
Phys.\ Rev.\ D {\bf 49}, 6857 (1994).
[hep-th/9402044].
}

\lref\ZupnikRY{
   B.~M.~Zupnik and D.~G.~Pak,
   ``Topologically Massive Gauge Theories In Superspace,''
Sov.\ Phys.\ J.\  {\bf 31}, 962 (1988).
}

\lref\IvanovFN{
   E.~A.~Ivanov,
   ``Chern-Simons matter systems with manifest N=2 supersymmetry,''
Phys.\ Lett.\ B {\bf 268}, 203 (1991).
}

\lref\NiemiRQ{
  A.~J.~Niemi and G.~W.~Semenoff,
  ``Axial Anomaly Induced Fermion Fractionization and Effective Gauge Theory Actions in Odd Dimensional Space-Times,''
Phys.\ Rev.\ Lett.\  {\bf 51}, 2077 (1983).
}

\lref\RedlichDV{
  A.~N.~Redlich,
  ``Parity Violation and Gauge Noninvariance of the Effective Gauge Field Action in Three-Dimensions,''
Phys.\ Rev.\ D {\bf 29}, 2366 (1984).
}

\lref\AharonyGP{
  O.~Aharony,
  ``IR duality in d = 3 N=2 supersymmetric USp(2N(c)) and U(N(c)) gauge theories,''
Phys.\ Lett.\ B {\bf 404}, 71 (1997).
[hep-th/9703215].
}

\lref\AffleckAS{
  I.~Affleck, J.~A.~Harvey and E.~Witten,
  ``Instantons and (Super)Symmetry Breaking in (2+1)-Dimensions,''
Nucl.\ Phys.\ B {\bf 206}, 413 (1982).
}

\lref\BeemMB{
  C.~Beem, T.~Dimofte and S.~Pasquetti,
  ``Holomorphic Blocks in Three Dimensions,''
[arXiv:1211.1986 [hep-th]].
}

\lref\HwangJH{
  C.~Hwang, H.~-C.~Kim and J.~Park,
  ``Factorization of the 3d superconformal index,''
[arXiv:1211.6023 [hep-th]].
}

\lref\KrattenthalerDA{
  C.~Krattenthaler, V.~P.~Spiridonov, G.~S.~Vartanov,
  ``Superconformal indices of three-dimensional theories related by mirror symmetry,''
JHEP {\bf 1106}, 008 (2011).
[arXiv:1103.4075 [hep-th]].
}

\lref\GaddeEN{
  A.~Gadde, L.~Rastelli, S.~S.~Razamat and W.~Yan,
  ``On the Superconformal Index of N=1 IR Fixed Points: A Holographic Check,''
JHEP {\bf 1103}, 041 (2011).
[arXiv:1011.5278 [hep-th]].
}

\lref\ImamuraWG{
  Y.~Imamura and D.~Yokoyama,
 ``N=2 supersymmetric theories on squashed three-sphere,''
Phys.\ Rev.\ D {\bf 85}, 025015 (2012).
[arXiv:1109.4734 [hep-th]].
}

\lref\deBoerKA{
  J.~de Boer, K.~Hori, Y.~Oz and Z.~Yin,
  ``Branes and mirror symmetry in N=2 supersymmetric gauge theories in three-dimensions,''
Nucl.\ Phys.\ B {\bf 502}, 107 (1997).
[hep-th/9702154].
}

\lref\ClossetRU{
  C.~Closset, T.~T.~Dumitrescu, G.~Festuccia and Z.~Komargodski,
  ``Supersymmetric Field Theories on Three-Manifolds,''
JHEP {\bf 1305}, 017 (2013).
[arXiv:1212.3388 [hep-th]].
}

\lref\ImamuraRQ{
  Y.~Imamura and D.~Yokoyama,
 ``$S^3/Z_n$ partition function and dualities,''
JHEP {\bf 1211}, 122 (2012).
[arXiv:1208.1404 [hep-th]].
}

\lref\KapustinSim{
A.~Kapustin,  2010 Simons Workshop talk, a video of this talk can be found at
{\tt
http://media.scgp.stonybrook.edu/video/video.php?f=20110810\_1\_qtp.mp4}
}

\lref\PolyakovFU{
  A.~M.~Polyakov,
  ``Quark Confinement and Topology of Gauge Groups,''
Nucl.\ Phys.\ B {\bf 120}, 429 (1977).
}

\lref\newIS{
K.~Intriligator and N.~Seiberg,
  ``Aspects of 3d N=2 Chern-Simons-Matter Theories,''
[arXiv:1305.1633 [hep-th]].
}

\lref\BhattacharyaZY{
  J.~Bhattacharya, S.~Bhattacharyya, S.~Minwalla and S.~Raju,
  ``Indices for Superconformal Field Theories in 3,5 and 6 Dimensions,''
JHEP {\bf 0802}, 064 (2008).
[arXiv:0801.1435 [hep-th]].
}

\lref\JafferisUN{
  D.~L.~Jafferis,
  ``The Exact Superconformal R-Symmetry Extremizes Z,''
JHEP {\bf 1205}, 159 (2012).
[arXiv:1012.3210 [hep-th]].
}

\lref\JafferisZI{
  D.~L.~Jafferis, I.~R.~Klebanov, S.~S.~Pufu and B.~R.~Safdi,
  ``Towards the F-Theorem: N=2 Field Theories on the Three-Sphere,''
JHEP {\bf 1106}, 102 (2011).
[arXiv:1103.1181 [hep-th]].
}

\lref\IntriligatorID{
  K.~A.~Intriligator and N.~Seiberg,
  ``Duality, monopoles, dyons, confinement and oblique confinement in supersymmetric SO(N(c)) gauge theories,''
Nucl.\ Phys.\ B {\bf 444}, 125 (1995).
[hep-th/9503179].
}

\lref\SeibergPQ{
  N.~Seiberg,
  ``Electric - magnetic duality in supersymmetric nonAbelian gauge theories,''
Nucl.\ Phys.\ B {\bf 435}, 129 (1995).
[hep-th/9411149].
}

\lref\StrasslerFE{
  M.~J.~Strassler,
  ``Duality, phases, spinors and monopoles in $SO(N)$ and $spin(N)$ gauge theories,''
JHEP {\bf 9809}, 017 (1998).
[hep-th/9709081].
}
\lref\AharonyBX{
  O.~Aharony, A.~Hanany, K.~A.~Intriligator, N.~Seiberg and M.~J.~Strassler,
  ``Aspects of N=2 supersymmetric gauge theories in three-dimensions,''
Nucl.\ Phys.\ B {\bf 499}, 67 (1997).
[hep-th/9703110].
}

\lref\IntriligatorNE{
  K.~A.~Intriligator and P.~Pouliot,
  ``Exact superpotentials, quantum vacua and duality in supersymmetric SP(N(c)) gauge theories,''
Phys.\ Lett.\ B {\bf 353}, 471 (1995).
[hep-th/9505006].
}

\lref\IntriligatorAU{
  K.~A.~Intriligator and N.~Seiberg,
  ``Lectures on supersymmetric gauge theories and electric - magnetic duality,''
Nucl.\ Phys.\ Proc.\ Suppl.\  {\bf 45BC}, 1 (1996).
[hep-th/9509066].
}

\lref\KarchUX{
  A.~Karch,
  ``Seiberg duality in three-dimensions,''
Phys.\ Lett.\ B {\bf 405}, 79 (1997).
[hep-th/9703172].
}
\lref\MaldacenaSS{
  J.~M.~Maldacena, G.~W.~Moore and N.~Seiberg,
  ``D-brane charges in five-brane backgrounds,''
JHEP {\bf 0110}, 005 (2001).
[hep-th/0108152].
}

\lref\BanksZN{
  T.~Banks and N.~Seiberg,
  ``Symmetries and Strings in Field Theory and Gravity,''
Phys.\ Rev.\ D {\bf 83}, 084019 (2011).
[arXiv:1011.5120 [hep-th]].
}

\lref\GaiottoBE{
  D.~Gaiotto, G.~W.~Moore and A.~Neitzke,
  ``Framed BPS States,''
[arXiv:1006.0146 [hep-th]].
}
\lref\SafdiRE{
  B.~R.~Safdi, I.~R.~Klebanov and J.~Lee,
  ``A Crack in the Conformal Window,''
[arXiv:1212.4502 [hep-th]].
}

\lref\KutasovNP{
  D.~Kutasov and A.~Schwimmer,
  ``On duality in supersymmetric Yang-Mills theory,''
Phys.\ Lett.\ B {\bf 354}, 315 (1995).
[hep-th/9505004].
}

\lref\KutasovSS{
  D.~Kutasov, A.~Schwimmer and N.~Seiberg,
  ``Chiral rings, singularity theory and electric - magnetic duality,''
Nucl.\ Phys.\ B {\bf 459}, 455 (1996).
[hep-th/9510222].
}

\lref\GiveonZN{
  A.~Giveon and D.~Kutasov,
  ``Seiberg Duality in Chern-Simons Theory,''
Nucl.\ Phys.\ B {\bf 812}, 1 (2009).
[arXiv:0808.0360 [hep-th]].
}

\lref\HoriDK{
  K.~Hori and D.~Tong,
  ``Aspects of Non-Abelian Gauge Dynamics in Two-Dimensional N=(2,2) Theories,''
JHEP {\bf 0705}, 079 (2007).
[hep-th/0609032].
}

\lref\NiarchosAH{
  V.~Niarchos,
  ``Seiberg dualities and the 3d/4d connection,''
JHEP {\bf 1207}, 075 (2012).
[arXiv:1205.2086 [hep-th]].
}

\lref\KapustinJM{
  A.~Kapustin and B.~Willett,
  ``Generalized Superconformal Index for Three Dimensional Field Theories,''
[arXiv:1106.2484 [hep-th]].
}

\lref\KimCMA{
  H.~Kim and J.~Park,
  ``Aharony Dualities for 3d Theories with Adjoint Matter,''
[arXiv:1302.3645 [hep-th]].
}

\lref\KapustinVZ{
  A.~Kapustin, H.~Kim and J.~Park,
  ``Dualities for 3d Theories with Tensor Matter,''
JHEP {\bf 1112}, 087 (2011).
[arXiv:1110.2547 [hep-th]].
}

\lref\AharonyGP{
  O.~Aharony,
  ``IR duality in d = 3 N=2 supersymmetric USp(2N(c)) and U(N(c)) gauge theories,''
Phys.\ Lett.\ B {\bf 404}, 71 (1997).
[hep-th/9703215].
}

\lref\WittenDS{
  E.~Witten,
  ``Supersymmetric index of three-dimensional gauge theory,''
In *Shifman, M.A. (ed.): The many faces of the superworld* 156-184.
[hep-th/9903005].
}

\lref\FestucciaWS{
  G.~Festuccia and N.~Seiberg,
  ``Rigid Supersymmetric Theories in Curved Superspace,''
JHEP {\bf 1106}, 114 (2011).
[arXiv:1105.0689 [hep-th]].
}

\lref\SpiridonovHF{
  V.~P.~Spiridonov and G.~S.~Vartanov,
  ``Elliptic hypergeometry of supersymmetric dualities II. Orthogonal groups, knots, and vortices,''
[arXiv:1107.5788 [hep-th]].
}
\lref\SpiridonovZR{
  V.~P.~Spiridonov and G.~S.~Vartanov,
  ``Superconformal indices for N = 1 theories with multiple duals,''
Nucl.\ Phys.\ B {\bf 824}, 192 (2010).
[arXiv:0811.1909 [hep-th]].
}

\lref\HoriPD{
  K.~Hori,
  ``Duality In Two-Dimensional (2,2) Supersymmetric Non-Abelian Gauge Theories,''
[arXiv:1104.2853 [hep-th]].
}

\lref\RomelsbergerEG{
  C.~Romelsberger,
  ``Counting chiral primaries in N = 1, d=4 superconformal field theories,''
Nucl.\ Phys.\ B {\bf 747}, 329 (2006).
[hep-th/0510060].
}

\lref\KapustinKZ{
  A.~Kapustin, B.~Willett and I.~Yaakov,
  ``Exact Results for Wilson Loops in Superconformal Chern-Simons Theories with Matter,''
JHEP {\bf 1003}, 089 (2010).
[arXiv:0909.4559 [hep-th]].
}

\lref\SpiridonovHF{
  V.~P.~Spiridonov and G.~S.~Vartanov,
  ``Elliptic hypergeometry of supersymmetric dualities II. Orthogonal groups, knots, and vortices,''
[arXiv:1107.5788 [hep-th]].
}

\lref\NaculichNC{
  S.~G.~Naculich and H.~J.~Schnitzer,
  ``Level-rank duality of the U(N) WZW model, Chern-Simons theory, and 2-D qYM theory,''
JHEP {\bf 0706}, 023 (2007).
[hep-th/0703089 [HEP-TH]].
}

\lref\Naculich{
  S.~G.~Naculich, H.~A.~Riggs and H.~J.~Schnitzer,
  ``Group Level Duality In Wzw Models And Chern-simons Theory,''
  Phys.\ Lett.\ B {\bf 246} (1990) 417.
}

\lref\RazamatUV{
  S.~S.~Razamat,
  ``On a modular property of N=2 superconformal theories in four dimensions,''
JHEP {\bf 1210}, 191 (2012).
[arXiv:1208.5056 [hep-th]].
}
\lref\Camperi{
  M.~Camperi, F.~Levstein and G.~Zemba,
  ``The Large N Limit Of Chern-simons Gauge Theory,''
  Phys.\ Lett.\ B {\bf 247} (1990) 549.
}

\lref\Mlawer{
  E.~J.~Mlawer, S.~G.~Naculich, H.~A.~Riggs and H.~J.~Schnitzer,
  ``Group level duality of WZW fusion coefficients and Chern-Simons link observables,''
  Nucl.\ Phys.\ B {\bf 352} (1991) 863.
}

\lref\DolanQI{
  F.~A.~Dolan and H.~Osborn,
  ``Applications of the Superconformal Index for Protected Operators and q-Hypergeometric Identities to N=1 Dual Theories,''
Nucl.\ Phys.\ B {\bf 818}, 137 (2009).
[arXiv:0801.4947 [hep-th]].
}

\lref\EagerHX{
  R.~Eager, J.~Schmude and Y.~Tachikawa,
  ``Superconformal Indices, Sasaki-Einstein Manifolds, and Cyclic Homologies,''
[arXiv:1207.0573 [hep-th]].
}

\lref\GaddeIA{
  A.~Gadde and W.~Yan,
  ``Reducing the 4d Index to the $S^3$ Partition Function,''
JHEP {\bf 1212}, 003 (2012).
[arXiv:1104.2592 [hep-th]].
}

\lref\DolanRP{
  F.~A.~H.~Dolan, V.~P.~Spiridonov and G.~S.~Vartanov,
  ``From 4d superconformal indices to 3d partition functions,''
Phys.\ Lett.\ B {\bf 704}, 234 (2011).
[arXiv:1104.1787 [hep-th]].
}

\lref\IntriligatorEX{
  K.~A.~Intriligator and N.~Seiberg,
  ``Mirror symmetry in three-dimensional gauge theories,''
Phys.\ Lett.\ B {\bf 387}, 513 (1996).
[hep-th/9607207].
}

\lref\deBoerMP{
  J.~de Boer, K.~Hori, H.~Ooguri and Y.~Oz,
  ``Mirror symmetry in three-dimensional gauge theories, quivers and D-branes,''
Nucl.\ Phys.\ B {\bf 493}, 101 (1997).
[hep-th/9611063].
}

\lref\ImamuraUW{
  Y.~Imamura,
 ``Relation between the 4d superconformal index and the $S^3$ partition function,''
JHEP {\bf 1109}, 133 (2011).
[arXiv:1104.4482 [hep-th]].
}

\lref\SeibergBZ{
  N.~Seiberg,
  ``Exact results on the space of vacua of four-dimensional SUSY gauge theories,''
Phys.\ Rev.\ D {\bf 49}, 6857 (1994).
[hep-th/9402044].
}

\lref\HamaEA{
  N.~Hama, K.~Hosomichi and S.~Lee,
  ``SUSY Gauge Theories on Squashed Three-Spheres,''
JHEP {\bf 1105}, 014 (2011).
[arXiv:1102.4716 [hep-th]].
}

\lref\AffleckAS{
  I.~Affleck, J.~A.~Harvey and E.~Witten,
  ``Instantons and (Super)Symmetry Breaking in (2+1)-Dimensions,''
Nucl.\ Phys.\ B {\bf 206}, 413 (1982).
}

\lref\SeibergPQ{
  N.~Seiberg,
  ``Electric - magnetic duality in supersymmetric nonAbelian gauge theories,''
Nucl.\ Phys.\ B {\bf 435}, 129 (1995).
[hep-th/9411149].
}

\lref\CveticXN{
  M.~Cvetic, T.~W.~Grimm and D.~Klevers,
  ``Anomaly Cancellation And Abelian Gauge Symmetries In F-theory,''
JHEP {\bf 1302}, 101 (2013).
[arXiv:1210.6034 [hep-th]].
}

\lref\debult{
  F.~van~de~Bult,
  ``Hyperbolic Hypergeometric Functions,''
University of Amsterdam Ph.D. thesis
}

\lref\MoritaCS{
  T.~Morita and V.~Niarchos,
  ``F-theorem, duality and SUSY breaking in one-adjoint Chern-Simons-Matter theories,''
Nucl.\ Phys.\ B {\bf 858}, 84 (2012).
[arXiv:1108.4963 [hep-th]].
}

\lref\Shamirthesis{
  I.~Shamir,
  ``Aspects of three dimensional Seiberg duality,''
  M. Sc. thesis submitted to the Weizmann Institute of Science, April 2010.
  }

\lref\slthreeZ{
  J.~Felder, A.~Varchenko,
  ``The elliptic gamma function and $SL(3,Z) \times Z^3$,'' $\;\;$
[arXiv:math/0001184].
}

\lref\SpiridonovZA{
  V.~P.~Spiridonov and G.~S.~Vartanov,
  ``Elliptic Hypergeometry of Supersymmetric Dualities,''
Commun.\ Math.\ Phys.\  {\bf 304}, 797 (2011).
[arXiv:0910.5944 [hep-th]].
}

\lref\ZwiebelWA{
  B.~I.~Zwiebel,
  ``Charging the Superconformal Index,''
JHEP {\bf 1201}, 116 (2012).
[arXiv:1111.1773 [hep-th]].
}

\lref\BeniniMF{
  F.~Benini, C.~Closset and S.~Cremonesi,
  ``Comments on 3d Seiberg-like dualities,''
JHEP {\bf 1110}, 075 (2011).
[arXiv:1108.5373 [hep-th]].
}

\lref\ClossetVG{
  C.~Closset, T.~T.~Dumitrescu, G.~Festuccia, Z.~Komargodski and N.~Seiberg,
  ``Contact Terms, Unitarity, and F-Maximization in Three-Dimensional Superconformal Theories,''
JHEP {\bf 1210}, 053 (2012).
[arXiv:1205.4142 [hep-th]].
}

\lref\ClossetVP{
  C.~Closset, T.~T.~Dumitrescu, G.~Festuccia, Z.~Komargodski and N.~Seiberg,
  ``Comments on Chern-Simons Contact Terms in Three Dimensions,''
JHEP {\bf 1209}, 091 (2012).
[arXiv:1206.5218 [hep-th]].
}

\lref\SpiridonovHF{
  V.~P.~Spiridonov and G.~S.~Vartanov,
  ``Elliptic hypergeometry of supersymmetric dualities II. Orthogonal groups, knots, and vortices,''
[arXiv:1107.5788 [hep-th]].
}

\lref\ElitzurFH{
  S.~Elitzur, A.~Giveon and D.~Kutasov,
  ``Branes and N=1 duality in string theory,''
Phys.\ Lett.\ B {\bf 400}, 269 (1997).
[hep-th/9702014].
}

\lref\ElitzurHC{
  S.~Elitzur, A.~Giveon, D.~Kutasov, E.~Rabinovici and A.~Schwimmer,
  ``Brane dynamics and N=1 supersymmetric gauge theory,''
Nucl.\ Phys.\ B {\bf 505}, 202 (1997).
[hep-th/9704104].
}

\lref\KapustinGH{
  A.~Kapustin,
  ``Seiberg-like duality in three dimensions for orthogonal gauge groups,''
[arXiv:1104.0466 [hep-th]].
}

\lref\HwangHT{
  C.~Hwang, K.~-J.~Park and J.~Park,
  ``Evidence for Aharony duality for orthogonal gauge groups,''
JHEP {\bf 1111}, 011 (2011).
[arXiv:1109.2828 [hep-th]].
}

\lref\SpiridonovWW{
  V.~P.~Spiridonov and G.~S.~Vartanov,
  ``Elliptic hypergeometric integrals and 't Hooft anomaly matching conditions,''
JHEP {\bf 1206}, 016 (2012).
[arXiv:1203.5677 [hep-th]].
}

\lref\DimoftePY{
  T.~Dimofte, D.~Gaiotto and S.~Gukov,
  ``3-Manifolds and 3d Indices,''
[arXiv:1112.5179 [hep-th]].
}

\lref\KimWB{
  S.~Kim,
  ``The Complete superconformal index for N=6 Chern-Simons theory,''
Nucl.\ Phys.\ B {\bf 821}, 241 (2009), [Erratum-ibid.\ B {\bf 864}, 884 (2012)].
[arXiv:0903.4172 [hep-th]].
}

\lref\WillettGP{
  B.~Willett and I.~Yaakov,
  ``N=2 Dualities and Z Extremization in Three Dimensions,''
[arXiv:1104.0487 [hep-th]].
}

\lref\KapustinXQ{
  A.~Kapustin, B.~Willett and I.~Yaakov,
  ``Nonperturbative Tests of Three-Dimensional Dualities,''
JHEP {\bf 1010}, 013 (2010).
[arXiv:1003.5694 [hep-th]].
}

\lref\ImamuraSU{
  Y.~Imamura and S.~Yokoyama,
  ``Index for three dimensional superconformal field theories with general R-charge assignments,''
JHEP {\bf 1104}, 007 (2011).
[arXiv:1101.0557 [hep-th]].
}

\lref\GaddeDDA{
  A.~Gadde and S.~Gukov,
  ``2d Index and Surface operators,''
[arXiv:1305.0266 [hep-th]].
}

\lref\HwangQT{
  C.~Hwang, H.~Kim, K.~-J.~Park and J.~Park,
  ``Index computation for 3d Chern-Simons matter theory: test of Seiberg-like duality,''
JHEP {\bf 1109}, 037 (2011).
[arXiv:1107.4942 [hep-th]].
}

\lref\GreenDA{
  D.~Green, Z.~Komargodski, N.~Seiberg, Y.~Tachikawa and B.~Wecht,
  ``Exactly Marginal Deformations and Global Symmetries,''
JHEP {\bf 1006}, 106 (2010).
[arXiv:1005.3546 [hep-th]].
}

\lref\IntriligatorID{
  K.~A.~Intriligator and N.~Seiberg,
  ``Duality, monopoles, dyons, confinement and oblique confinement in supersymmetric SO(N(c)) gauge theories,''
Nucl.\ Phys.\ B {\bf 444}, 125 (1995).
[hep-th/9503179].
}

\lref\SeibergQD{
  N.~Seiberg,
  ``Modifying the Sum Over Topological Sectors and Constraints on Supergravity,''
JHEP {\bf 1007}, 070 (2010).
[arXiv:1005.0002 [hep-th]].
}
\lref\BanksZN{
  T.~Banks and N.~Seiberg,
  ``Symmetries and Strings in Field Theory and Gravity,''
Phys.\ Rev.\ D {\bf 83}, 084019 (2011).
[arXiv:1011.5120 [hep-th]].
}

\lref\SeibergNZ{
  N.~Seiberg and E.~Witten,
  ``Gauge dynamics and compactification to three-dimensions,''
In *Saclay 1996, The mathematical beauty of physics* 333-366.
[hep-th/9607163].
}

\lref\AharonyCI{
  O.~Aharony and I.~Shamir,
  ``On $O(N_c)$ d=3 N=2 supersymmetric QCD Theories,''
JHEP {\bf 1112}, 043 (2011).
[arXiv:1109.5081 [hep-th]].
}

\lref\GiveonSR{
  A.~Giveon and D.~Kutasov,
  ``Brane dynamics and gauge theory,''
Rev.\ Mod.\ Phys.\  {\bf 71}, 983 (1999).
[hep-th/9802067].
}

\lref\KapustinPY{
  A.~Kapustin,
  ``Wilson-'t Hooft operators in four-dimensional gauge theories and S-duality,''
Phys.\ Rev.\ D {\bf 74}, 025005 (2006).
[hep-th/0501015].
}

\lref\KapustinGH{
  A.~Kapustin,
  ``Seiberg-like duality in three dimensions for orthogonal gauge groups,''
[arXiv:1104.0466 [hep-th]].
}

\lref\ARSW{
  O.~Aharony, S.~S.~Razamat, N.~Seiberg and B.~Willett,
  ``3d dualities from 4d dualities,''
[arXiv:1305.3924 [hep-th]].
}

\lref\AST{
O.~Aharony, N.~Seiberg and Y.~Tachikawa,
  ``Reading between the lines of four-dimensional gauge theories,''
[arXiv:1305.0318 [hep-th]].
}

\Title{\vbox{\baselineskip12pt
\hbox{WIS/07/13-JUN-DPPA}
}}
{\vbox{\centerline{$3d$ dualities from $4d$ dualities for orthogonal groups}
}}
\centerline{Ofer Aharony$^{a,b}$, Shlomo S. Razamat$^a$, Nathan Seiberg$^a$,
and Brian Willett$^a$}
\bigskip
\centerline{$^a${\it School of Natural Sciences, Institute for Advanced Study, Princeton, NJ 08540, USA}}
\centerline{}
\centerline{${}^b${\it Department of Particle Physics and Astrophysics}}
\centerline{{\it Weizmann Institute of Science, Rehovot 76100, Israel}}
\vskip.1in \vskip.1in \centerline{\bf Abstract}

\noindent
We extend recent work on the relation of $4d$ and $3d$ IR dualities of supersymmetric gauge theories with four supercharges to the case of orthogonal gauge groups. The distinction between different $SO(N)$ gauge theories in $4d$ plays an important role in this relation. We show that the $4d$ duality leads to a $3d$ duality between an $SO(N_c)$ gauge theory with $N_f$ flavors and an $SO(N_f-N_c+2)$ theory with $N_f$ flavors and extra singlets, and we derive its generalization in the presence of Chern-Simons terms. 
There are two different $O(N)$ theories in $3d$, which we denote by $O(N)_\pm$, and we also show that the $O(N_c)_-$ gauge theory is dual to a $Spin(N_f-N_c+2)$ theory, and derive from $4d$ the known duality between $O(N_c)_+$ and $O(N_f-N_c+2)_+$. We verify the consistency of these $3d$ dualities by various methods, including index computations.

\vfill

\Date{July 2013}


\newsec{Introduction}

A crucial role in gauge theory dynamics is played by various dualities. They give a weakly coupled description of some strong coupling phenomena (like confinement and chiral symmetry breaking), and may point to a deep structure underlying the theory.  In different situations these dualities manifest themselves differently.  Some $4d$ superconformal theories like $\NN=4$ and certain $\NN=2$ supersymmetric theories exhibit exact electric/magnetic duality, leading to several distinct descriptions of the same theory, with different values of the coupling constant and sometimes even different gauge groups. Many four-dimensional $\NN=1,\ 2$, three-dimensional $\NN=2,\ 3,\cdots$, and certain two-dimensional theories have IR dualities, relating different theories with the same IR limit \SeibergPQ. In some situations, including $4d$ $\NN=1$ $SO(N)$ dualities, it is clear that these are also related to electric/magnetic duality \refs{\SeibergPQ,\IntriligatorID,\IntriligatorAU}; when the gauge group is broken to $SO(2)$ they reduce to an Abelian electric/magnetic duality, and they exchange Wilson lines with 't~Hooft-Wilson lines \refs{\StrasslerFE,\AST}.

In $3d$ there are several known examples of such IR dualities, both with $\NN=2$ and with higher supersymmetries.  In a previous paper we argued that most and perhaps all such dualities in $3d$ originate from ancestor dualities in $4d$ \ARSW\ (see also \NiarchosAH).  The purpose of this note is to extend this discussion to theories with orthogonal gauge groups.

The discussion in \ARSW\ starts with any $4d$ $\NN=1$ duality, and by carefully compactifying it on a circle, it leads to a clear prescription for how to generate from it a corresponding $3d$ duality. For example, we can start with the characteristic example of a $4d$ $\NN=1$ duality.  This is the duality between an $SU(N_c)$ gauge theory with $N_f$ flavors $Q^i$ and $\tilde Q_{\tilde i}$,  and its dual $SU(N_f-N_c)$ gauge theory with $N_f$ dual quarks $q_i$ and $\tilde q^{\tilde i}$  and elementary gauge neutral ``mesons'' $M_{\tilde i}^i$ and a superpotential $W=M_{\tilde i}^i q_i \tilde q^{\tilde i}$ \SeibergPQ.  It is common to refer to these theories as the electric and the magnetic theories, but we will refer to them as theory A and theory B.  A naive dimensional reduction of any of these two dual theories to $3d$ leads to a theory with an additional ``axial'' $U(1)$ global symmetry. This is the symmetry that is anomalous in $4d$, but is preserved in $3d$.  The prescription of \ARSW\ is to modify the naive dimensionally reduced theory by adding to its Lagrangian a suitable operator, generated by non-perturbative effects in the theory on a circle, which explicitly breaks this anomalous $U(1)$ symmetry.  In theory A we add a superpotential
\eqn\WAeta{W_A= \eta Y ~,}
where $\eta = \Lambda^{b_0}$ is the instanton factor \IntriligatorAU\ of theory A, and $Y$ is its monopole operator.  In theory B, which already had a superpotential in $4d$, we have
\eqn\WBeta{W_B= M_{\tilde i}^i q_i \tilde q^{\tilde i}+\tilde \eta \tilde Y~,}
where $\tilde \eta = \tilde \Lambda^{\tilde b_0}= (-1)^{N_f-N_c} \eta^{-1}$ is the instanton factor of theory B, and $\tilde Y$ is its monopole operator.\foot{The first term in \WBeta\ contains already in $4d$ a
scale $\mu$ \IntriligatorAU, which is related to the normalization of the kinetic terms of $M$, $q$ and $\tilde q$.  It is natural to normalize $M$ such that it is identified with the composite operator $M=Q {\tilde Q}$ in theory A.  We will choose this normalization.
When we reduce to $3d$ the parameter $\mu$ is still present, and there are also various factors of the radius. Here we ignore this normalization, which is irrelevant in the IR.  We fix arbitrarily (but self-consistently) the coefficient of the first term in \WBeta\ ($\mu=1$) and in the analogous expressions for $SO(N_c)$.}
The arguments of \ARSW\ imply that the two $3d$ theories \WAeta, \WBeta\ are equivalent at low energies.

Once such a $3d$ duality is established one can find many additional $3d$ dualities, which follow from it.  First, we can turn on relevant operators in the two sides of the duality and flow to the IR.  Second, we can gauge any of the global symmetries of the theories and generate new dual pairs.  These two tools were used in \ARSW\ to reproduce all the known dualities between $3d$ $\NN=2$ theories with $SU(N_c)$, $U(N_c)$ and $USp(2N_c)$ gauge groups, and to generate many new dualities.

However, the application of this procedure to theories with orthogonal gauge groups turns out to need more care.  In fact, already in $4d$ $\NN=1$ theories the IR dualities for orthogonal groups are significantly more subtle than for unitary or symplectic gauge groups \refs{\SeibergPQ,\IntriligatorID,\IntriligatorAU,\IntriligatorER}.  One underlying reason for this complexity was recently identified in \AST.  It is known that if the Lie algebra of the gauge symmetry is $so(N_c)$, the gauge group can be $Spin(N_c)$ or $SO(N_c)$ (and it could even have disconnected components, making it $Pin(N_c)$ or $O(N_c)$). The main point of \AST\ is that even when the gauge group is $SO(N_c)$, there are two distinct $4d$ gauge theories with that gauge group, denoted by $SO(N_c)_\pm$. In the Euclidean path integral they are distinguished by a new term in the Lagrangian -- a certain $\Z_2$-valued theta-like-angle, associated with the Pontryagin square $\PP(w_2)$ of the Steifel-Whitney class $w_2$ of the gauge bundle.

A simple physical way to distinguish between the three gauge theories $SO(N_c)_\pm$ and $Spin(N_c)$ already in $\R^4$ is to study their line operators.  The $Spin(N_c)$ theory has a Wilson loop $W$ in a spinor representation\foot{In this note we will study $so(N_c)$ gauge theories with matter fields in the vector representation.  Therefore, for many purposes we can identify Wilson loops in different spinor representations.  For a more detailed discussion see \AST.}.  Its square $W^2$ can be screened by dynamical fields and we will view it as trivial.  The two $SO(N_c)$ theories do not have a Wilson loop in a spinor representation.  Instead, they have 't~Hooft loops carrying smaller magnetic charge than is allowed in $Spin(N_c)$.  The $SO(N_c)_+$ theory has a purely magnetic 't~Hooft loop operator $H$, and the $SO(N_c)_-$ has the non-trivial loop operator $HW$.  For a closely related earlier discussion, see \GaiottoBE.

For $N_c=3$ the distinction between $SO(3)_\pm$ can be understood by extending the range of the ordinary theta-angle to be in $[0,4\pi)$, and then \GaiottoBE
\eqn\sothree{SO(3)_+^\theta  = SO(3)_-^{\theta + 2\pi}.}
Similarly, for $N_c=4$ we have $Spin(4) =SU(2) \times SU(2)$ and $SO(4) = (SU(2) \times SU(2) /\Z_2$.  Hence, the $so(4)$ theory has two theta-angles, one for each $SU(2)$, and
\eqn\sofour{SO(4)_+^{\theta_1,\theta_2}  = SO(4)_-^{\theta_1, \theta_2 + 2\pi} = SO(4)_-^{\theta_1 + 2\pi, \theta_2}= SO(4)_+^{\theta_1 + 2\pi, \theta_2+ 2\pi}  .}
For higher values of $N_c$ the distinction between the two $SO(N_c)$ theories cannot be absorbed in extending the range of $\theta$, and the only way to describe it at the level of the Lagrangian is by using the Pontryagin square $\PP(w_2)$.

The $4d$ $\NN=1$ IR duality relates \refs{\SeibergPQ,\IntriligatorID,\IntriligatorAU,\IntriligatorER} an $so(N_c)$ gauge theory with $N_f$ vectors $Q^i$ to an $so(\tilde N_c= N_f-N_c+4)$ gauge theory, with $N_f$ vectors $q_i$ and elementary gauge singlet mesons $M^{ij}$, and with a superpotential $W={1\over 2}M^{ij}q_iq_j$.  When either $N_c$ or $\tilde N_c$ are $2,\ 3,\ 4$ a more careful discussion is needed.  The analysis of \AST, which took into account the global structure\foot{For an earlier related discussion see \StrasslerFE.}, identified this duality as
\eqn\sodu{\eqalign{
&Spin(N_c) \to SO(\tilde N_c)_- ~,\cr
&SO(N_c)_- \to Spin(\tilde N_c) ~,\cr
&SO(N_c)_+ \to SO(\tilde N_c)_+ ~.}}

Subtleties associated with the line operators in the $4d$ theory translate into subtleties with the local operators when the theory is compactified on a circle to $3d$.  In particular, a $4d$ 't Hooft line operator $H$ wrapping this circle turns into a local monopole operator $Y$ in $3d$.  Hence, the choice of line operators in $4d$ becomes a choice of local operators in $3d$, which has more dramatic consequences, as we will see in our discussion below.

An additional subtlety in the analysis of orthogonal groups is that the corresponding $4d$ supersymmetric QCD (SQCD) theories on $\S^1$ have a Coulomb branch that is not lifted by quantum corrections. This did not occur in any of the cases analyzed in \ARSW, and the mapping of the Coulomb branch across the duality turns out to be non-trivial.

In section 2 we discuss some classical and quantum properties of $3d$ $\NN=2$ theories with orthogonal gauge groups.  We identify the coordinates on their moduli space of vacua, paying particular attention to the global structure (the distinction between $Spin(N)$ and $SO(N)$). In section 3 we discuss the $4d$ gauge theories on $\R^3\times \S^1$ and their moduli space of vacua.  Here the distinction between the three different theories with the same Lie algebra $so(N_c)$ is crucial.  In section 4 we follow \ARSW\ and consider two dual $4d$ theories compactified on a circle, and carefully identify their moduli spaces.

In  section 5 we derive the main result of this paper.  By taking an appropriate limit of the $4d$ theory on a circle we derive $3d$ dualities.  In particular, the $SO(N_c)$ SQCD theory with $N_f$ vectors $Q$ is dual to an $SO(N_f-N_c+2)$ gauge theory with $N_f$ vectors $q$, with gauge singlet fields $M$ and $Y$, and with an appropriate superpotential.  Here the elementary fields $M$ and $Y$ are identified with the composite meson $QQ$ and monopole operator of the original $SO(N_c)$ theory. This is closely related to dualities of $O(N_c)$ SQCD theories, previously found in \refs{\BeniniMF,\HwangHT,\AharonyCI}. 
We also find that the $Spin(N_c)$ theory is dual to an $O(N_f-N_c+2)_-$ theory, where $O(N)_-$ is a novel $3d$ $O(N)$ theory that we introduce in section 2.4. We perform various tests of these dualities, and deform them to obtain dualities for theories with Chern-Simons terms. Additional detailed tests are performed in section 6, where we discuss the $\S^2 \times \S^1$ and the $\S^3$ partition functions of all these theories.

\newsec{Background}

Much of the necessary background for this paper is found in the preceding paper \ARSW, and in
references therein. We will assume here familiarity with that paper, and discuss only the new
issues which arise for orthogonal gauge groups. Some aspects of the theory depend on the
precise choice of gauge group, while others depend only on the gauge algebra $g=so(N_c)$, and
we will try to distinguish the two in the following.

\subsec{Monopole operators and Coulomb branch coordinates for $g=so(N_c)$}

Three dimensional ${\cal N}=2$ gauge theories have classical Coulomb branches, where
the adjoint scalar $\sigma$ in the vector multiplet gets an expectation value, generically breaking
the gauge group $G$ to $U(1)^{r_G}$ (where $r_G$ is the rank of $G$). On this branch
we can dualize the $r_G$ photons to scalars $a_i$, or supersymmetrically dualize the $r_G$
$U(1)$ vector multiplets to chiral multiplets $Y_i$. The expectation values of these
chiral multiplets label the classical Coulomb branch of the theory. The chiral
multiplets $Y_i$ are ``monopole operators'' in the effective low-energy theory, creating a
$U(1)^{r_G}$ magnetic flux around them. In some cases they arise as low-energy limits of
microscopic ``monopole operators''. The allowed spectrum of monopole operators, and thus
the appropriate coordinates on the Coulomb branch, depends on the choice of the gauge
group; this choice determines the allowed Wilson line operators, and the monopole operators
need to be mutually local with respect to these Wilson lines.

For theories based
on the Lie algebra $g=so(N_c)$, when $N_c$ is even, $N_c=2r_G$ and when $N_c$ is odd,
$N_c=2r_G+1$. We write the adjoint matrix $\sigma$ as a matrix
in the vector representation of $so(N_c)$, and we can always diagonalize it. For
every non-zero eigenvalue, there is another eigenvalue of equal magnitude and opposite sign.
For even values of $N_c$ we write the eigenvalues as $\{\sigma_1,\cdots,\sigma_{r_G},-\sigma_{r_G},\cdots,
-\sigma_1\}$. By a Weyl transformation we can always choose
\eqn\ordering{N_c \ {\rm even}: \qquad \sigma_1 \geq \sigma_2 \geq \cdots \geq \sigma_{r_G-1} \geq |\sigma_{r_G}|.}
If our gauge group includes reflections (namely, it is $G=O(N_c)$ or $G=Pin(N_c)$ rather
than $G=SO(N_c)$ or $G=Spin(N_c)$) then we can also set $\sigma_{r_G} \geq 0$, while
otherwise we cannot do this in general. For odd values of $N_c$ we can write the
eigenvalues of $\sigma$ as $\{\sigma_1,\cdots,\sigma_{r_G},0,-\sigma_{r_G},\cdots,
-\sigma_1\}$, and by a Weyl transformation we can always choose
\eqn\orderingtwo{N_c \ {\rm odd}: \qquad \sigma_1 \geq \sigma_2 \geq \cdots \geq \sigma_{r_G-1} \geq \sigma_{r_G} \geq 0.}

The magnetic charges carried by the Coulomb branch coordinates should be thought of
as charges in the magnetic-dual algebra to $so(N_c)$. For even $N_c$, this algebra is
$so(N_c)$, and for odd $N_c$, it is $usp(N_c-1)$. There are always operators carrying
the charges of the roots of this algebra, and when the gauge group is $G=Spin(N_c)$,
these are the only allowed charges. We can then write the Coulomb branch coordinates
semi-classically as
\eqn\ccdef{
\eqalign{
Y_i \approx & \, \exp\left({{\sigma_i-\sigma_{i+1}}\over {\hat g}_3^2} + i (a_i - a_{i+1})\right),\quad
(i=1,\cdots,r_G-1),\cr
Y_{r_G} \approx & \cases{
\exp\left({{\sigma_{r_G-1}+\sigma_{r_G}}\over {\hat g}_3^2} + i (a_{r_G-1} + a_{r_G})\right) & for even $N_c$,\cr
& \cr
\exp\left({{2\sigma_{r_G}}\over {\hat g}_3^2} + i (2a_{r_G})\right) & for odd $N_c$. \cr}
}
}
Here ${\hat g}_3^2 = g_3^2 / 4\pi$, where $g_3$ is the gauge coupling constant of the $3d$ gauge theory, normalized as in \ARSW. The dependence of these operators on the $\sigma$'s that we wrote is
valid far out on the Coulomb branch, and gets quantum corrections, while their dependence on the dual
photons $a_i$ is exact.
As usual, the global symmetry charges of these operators can be determined by summing
over the charges of the fermions in chiral and vector multiplets,
which are coupled to the corresponding $\sigma$'s \AharonyBX. The $SO(N_c)$ and $Spin(N_c)$ theories have a global charge conjugation symmetry $\Z_2^{\cal C}$, which is gauged in the $O(N_c)$ and $Pin(N_c)$ theories. In the theories with even $N_c$, this symmetry exchanges the Coulomb branch coordinates $Y_{r_G-1}$ and $Y_{r_G}$.

As discussed in \refs{\newIS,\ARSW}, some of the Coulomb branch coordinates are low-energy
limits of microscopic monopole operators. These are defined so that their insertion at a
point $x$ generates some magnetic flux on the $\S^2$ surrounding $x$, and takes the $\sigma(y)$'s
pointing in the direction of the flux to $+\infty$ as $y\to x$. In the $Spin(N_c)$ theory, the monopole
operators all carry charges corresponding to roots of the dual magnetic group. The minimal
monopole operator $Y_{Spin}$ turns on one unit of flux, breaking $so(N_c) \to so(N_c-2)\times u(1)$, and
takes one of the eigenvalues of $\sigma$ to $\infty$. On the moduli space at low energies,
using \ordering\ and \orderingtwo, this monopole $Y_{Spin}$ looks semi-classically like
\eqn\ydef{Y_{Spin} \approx \exp\left({{2 \sigma_1}\over {\hat g}_3^2} + 2 i a_1\right)~.}
It is a combination of the Coulomb branch coordinates described in \ccdef,
\eqn\yspindecomp{Y_{Spin} = \cases{Y_1^2 Y_2^2 \cdots Y_{r_G-2}^2 Y_{r_G-1} Y_{r_G} & for even $N_c$,\cr
& \cr
Y_1^2 Y_2^2 \cdots Y_{r_G-1}^2 Y_{r_G} & for odd $N_c$.\cr}}

Another monopole operator that will play a role in our discussion is the one that takes
two eigenvalues of $\sigma$ to infinity together, breaking $so(N_c)\to so(N_c-4)\times u(2)$.
This monopole semi-classically looks like
\eqn\zdef{Z \approx \exp\left({{\sigma_1+\sigma_2}\over {\hat g}_3^2} + i (a_1 + a_2)\right),}
and we will see that it will play an important role in the discussion of $4d$ $so(N_c)$ theories
on a circle. It obeys $Y_{Spin} = Y_1 Z$, and
\eqn\zdecomp{Z = \cases{Y_1 Y_2^2 Y_3^2 \cdots Y_{r_G-2}^2 Y_{r_G-1} Y_{r_G} & for even $N_c$,\cr
& \cr
Y_1 Y_2^2 Y_3^2 \cdots Y_{r_G-1}^2 Y_{r_G} & for odd $N_c$.\cr}}

For $G=SO(N_c)$, Wilson lines carrying spinor charge are not allowed (we will always assume
that Wilson lines in the vector representation are allowed, since we will be interested
here in theories with matter fields in the vector representation). This means that
extra Coulomb branch coordinates and monopole operators are allowed, carrying weights which
are not roots of the dual magnetic group. For even $N_c$ they are allowed to carry weights
in the vector representation of the dual $so(N_c)$, and for odd $N_c$ in the
fundamental representation of the dual $usp(N_c-1)$. The basic monopole operator in this case
behaves semi-classically as \refs{\BeniniMF,\AharonyCI}
\eqn\ysodef{Y \approx \exp\left({{\sigma_1}\over {\hat g}_3^2} + i a_1\right),}
and it obeys
\eqn\yzrel{Y^2 = Y_{Spin} = Y_1 Z.}
All other ``new'' monopole operators that exist in this
case may be written as products of $Y$ with the operators corresponding to roots of the dual
gauge group. Note that while for the monopole operators corresponding to roots, there is
a classical 't~Hooft-Polyakov monopole solution (which is an instanton of the $3d$ theory)
that is associated with them, there is no such solution for $Y$ of \ysodef. But this is not
related to the definition of this operator, both microscopically and in the low-energy effective action.

The fact that the quantum numbers of the Coulomb branch coordinates are determined by those of the matter fields implies that they change when some matter fields go to infinite mass and decouple. In such cases we have a relation between the Coulomb branch coordinates in the high-energy and in the low-energy theories, which can usually be uniquely determined by matching their quantum numbers. For example, if we start from a $Spin(N_c)$ theory with $N_f$ chiral superfields in the vector representation, and give a mass $m$ to one of them in the superpotential, we have a relation of the form
\eqn\massrel{(Y_{Spin})_{N_c,N_f} = m (Y_{Spin})_{N_c,N_f-1}.}
Similarly, if we start from such a theory and break the gauge group to $Spin(N_c-1)$ with $(N_f-1)$ flavors, by giving an expectation value to one of the chiral superfields $Q$ in the vector representation, we have  a relation
\eqn\vevrel{(Y_{Spin})_{N_c,N_f} \vev{Q^2} = (Y_{Spin})_{N_c-1,N_f-1}.}

For low values of $N_c$, $N_c < 5$, some modifications are needed in our discussion.

For $N_c=2$ the gauge group
$G=Spin(2) = U(1)$, and instead of the operator $Y_{Spin}$ we have the two standard $U(1)$
Coulomb branch coordinates \AharonyBX
\eqn\spintwocoords{{\hat V}_+ \approx \exp\left({{2 \sigma}\over {\hat g}_3^2} + 2 i a\right),\qquad
{\hat V}_- \approx \exp\left(-{{2 \sigma}\over {\hat g}_3^2} - 2 i a\right),}
parameterizing the parts of the Coulomb branch with $\sigma$ positive and negative,
respectively. Note that in $Spin(2)$ we have particles of charge $\pm 1/2$ under the $U(1)$
group, and hence the normalization of the monopole operators is twice the usual normalization.
For $G=SO(2)$ all particles have integer $U(1)$ charge, and we have monopole operators
carrying half the charge of \spintwocoords, given by
\eqn\sotwocoords{V_+ \approx \exp\left({{\sigma}\over {\hat g}_3^2} + i a\right),\qquad
V_- \approx \exp\left(-{{\sigma}\over {\hat g}_3^2} - i a\right).}
The charge conjugation symmetry $\Z_2^{\cal C}$ exchanges
 ${\hat V}_+$ and ${\hat V}_-$ (or $V_+$ and $V_-$), and we will find it convenient to define
the linear combinations
\eqn\wpmdef{W_{\pm} \equiv V_+ \pm V_- ~,\qquad\qquad
\hat W_{\pm} \equiv \hat V_+ \pm \hat V_- ~,}
that are even and odd under $\Z_2^{\cal C}$.

For $N_c=3$, we have $Spin(3)=SU(2)$, and the monopole $Y_{Spin}$ that we defined above is
the standard Coulomb branch coordinate / monopole operator of the $SU(2)$ theory (usually denoted by $Y$ \AharonyBX). When the
gauge group is $SO(3)=SU(2)/\Z_2$, there exists a monopole operator of lower charge, which
we denoted by $Y$ above.

For $N_c=4$, the group $Spin(4)$ is equivalent to $SU(2)\times SU(2)$, and we can then have
separate Coulomb branch coordinates and monopole operators in the two $SU(2)$ factors. The
Coulomb branch coordinates of the two $SU(2)$'s correspond to $\sigma_1 \pm \sigma_2$ in our
notations above. Thus, the Coulomb branch coordinates of the two $SU(2)$'s, which we will denote
by $Y^{(1)}$ and $Y^{(2)}$, look semi-classically like the operators $Y_1$ and $Z$ discussed above. The operator $Y_{Spin}$
in this case is the product of these two $SU(2)$ operators. When the gauge group is $G = SO(4) = (SU(2)\times SU(2))/\Z_2$, one does not allow Wilson loops carrying a charge under the center
of each $SU(2)$ separately, but only under both $SU(2)$'s. In this case the operator $Y$ that
we defined in \ysodef\ exists, and squares to the product of the two $SU(2)$ monopole operators, $Y^2 = Y^{(1)} Y^{(2)}$. The charge conjugation symmetry $\Z_2^{\cal C}$
exchanges the two $SU(2)$ factors, exchanging $Y^{(1)}$ with $Y^{(2)}$. As in \wpmdef, it is convenient to define the combinations
\eqn\ypmdef{Y_\pm \equiv Y^{(1)} \pm Y^{(2)},}
which are even and odd under $\Z_2^{\cal C}$.

\subsec{The quantum moduli space of $3d$ ${\cal N}=2$ theories with $g=so(N_c)$}

In the quantum theory, most of the Coulomb branch described above is lifted. Whenever two
of the eigenvalues of $\sigma$ come together at a non-zero value, the corresponding $U(1)^2$
symmetry is enhanced to $U(2)$. As shown in \AffleckAS, the corresponding 't Hooft Polyakov
monopole-instanton solutions generate a superpotential in this case, which drives the
eigenvalues apart. For even values of $N_c$ we have this effect whenever $\sigma_i$ approaches
$\sigma_{i+1}$ for $i=1,2,\cdots,r_G-1$, and also when $\sigma_{r_G-1}$ approaches $-\sigma_{r_G}$.  Thus, in the $3d$ pure $g=so(N_c)$ theory with $N_c$ even we have an effective quantum superpotential
\eqn\quantumeven{N_c\ {\rm even}: \qquad W = \sum_{i=1}^{r_G-1} {1\over Y_i} + {1\over Y_{r_G}}~,}
which completely lifts the Coulomb branch. For odd values of $N_c$ there is always one eigenvalue
at $\sigma=0$, and when $\sigma_{r_G}$ approaches $0$, the corresponding $u(1)$ is enhanced to $so(3)$ from $3$ eigenvalues coming together at $\sigma=0$. There is a similar superpotential arising here, with a
different normalization \DaviesNW, so that
\eqn\quantumodd{N_c\ {\rm odd}: \qquad W = \sum_{i=1}^{r_G-1} {1\over Y_i} + {2\over Y_{r_G}}~.}
Again this completely lifts the Coulomb branch, so that the pure $3d$ supersymmetric Yang-Mills (SYM) theory has a runaway with no
supersymmetric vacua.

We can follow the reasoning used in \ARSW\ for $SU(N)$ gauge theories to show that \quantumeven\ and \quantumodd\ are in fact exact.  More precisely, they are exact as functions of the chiral superfields $Y_i$, but the $Y_i$ are complicated functions of $\sigma_i$ and $a_i$.

Similar superpotentials lift the Coulomb branch also for $g=so(3)=su(2)$, and for $g=so(4)=su(2)\oplus su(2)$, where we have a separate superpotential of this type in each $su(2)$ factor. For $g=so(2)=u(1)$ there is no such effect, and the Coulomb branch of the (free) pure gauge theory is simply a cylinder, labeled
by ${\hat V}_+ = 1 / {\hat V}_-$ for $G=Spin(2)$, and by $V_+ = 1 / V_-$ for $G=SO(2)$.

In theories with flavors in the vector representation that have no real mass, these flavors become massless whenever some eigenvalues of $\sigma$ vanish. This gives extra zero modes to monopole-instantons corresponding to eigenvalues coming together at $\sigma=0$, such that they no longer generate superpotentials; for odd $N_c$ this happens for the monopole $Y_{r_G}$, and for even $N_c$ it happens either for $Y_{r_G-1}$ or for $Y_{r_G}$, depending on the sign of $\sigma_{r_G}$. In these theories the moduli space is not completely lifted, but a one (complex) dimensional branch remains, where only a
single eigenvalue $\sigma_1$ is turned on (and $\sigma_2=\sigma_3=\cdots=\sigma_{r_G}=0$). In the $Spin(N_c)$ theories, this quantum Coulomb branch
may be parameterized by the operator $Y_{Spin}$ of \ydef, and for $G=SO(N_c)$ theories it can be
parameterized by $Y$ of \ysodef. Note that because of the quantum superpotentials \quantumeven, \quantumodd, the operator $Z$ is not a chiral operator in the low-energy $3d$ theory\foot{$Z$ is proportional to two supercharges acting on $(Y_{Spin} Y_1 K_{Y_1})$, where $K_{Y_1}$ is the derivative of the K\"ahler potential with respect to $Y_1$.}, but $Y$ and $Y_{Spin}$ still are. As discussed in \refs{\BeniniMF,\AharonyCI}, for low numbers of flavors, $N_f < N_c-2$, there are additional quantum effects that lift the Coulomb branch, while for $N_f \geq N_c-2$ the Coulomb branch is not lifted. For $N_f \geq N_c-1$, the quantum moduli space is the same as the classical moduli space (the discussion in \refs{\BeniniMF,\AharonyCI} is just for $G=O(N_c)$, but most of it can be generalized also to the other gauge groups with $g=so(N_c)$).

The $3d$ SQCD theory with $N_f$ chiral multiplets $Q_i$ in the vector representation has a global $SU(N_f)\times U(1)_A\times U(1)_R$ symmetry. We can choose the flavors $Q_i$ to transform in the fundamental of $SU(N_f)$, with one unit of $U(1)_A$ charge and no $U(1)_R$ charge. In this case, for $G=SO(N_c)$ theories the Coulomb branch coordinate $Y$ is a singlet of $SU(N_f)$, with $(-N_f)$ units of $U(1)_A$ charge and $(N_f-N_c+2)$ units of $U(1)_R$ charge. For $G=Spin(N_c)$ the operator $Y_{Spin}$ has the same charges as $Y^2$ in the $G=SO(N_c)$ theories.

Note that for $N_c=2$ our theories with flavors are the same as the $U(1)$ theories with flavors discussed
in \AharonyBX. However, for $N_c=3$ our $so(3)$ theories have matter in the triplet (adjoint) representation, so they are not the same as $su(2)$ theories with fundamental matter. In particular, for $N_c=3$ and $N_f=1$, the SQCD theory has enhanced supersymmetry, and it is the same as the $3d$ ${\cal N}=4$ SYM theory discussed in \SeibergNZ. For $N_c=4$ the matter fields are charged under both $SU(2)$'s, and couple them together. Depending on where we are in the moduli space, the superpotential involving the Coulomb branch coordinate of one of the $SU(2)$'s is lifted by the matter zero modes, while the other one remains. For $N_c=3,4$ we can still parameterize the remaining part of the Coulomb branch by $Y_{Spin}$ or $Y$, as for higher values of $N_c$.

\subsec{Baryon-monopole operators}
\subseclab\baryonmon

In the $g=so(N_c)$ SQCD theory with $N_f$ flavors $Q_i$, the list of chiral multiplets includes the monopole
operators discussed above, the mesons $M_{ij} = Q_i Q_j$ (symmetric in $i,j$) and (for $N_f \geq N_c$, and for $G=SO(N_c)$ or $G=Spin(N_c)$) the baryons $B = Q^{N_c}$, contracted with an epsilon symbol
in $so(N_c)$.  The operator $B^2$ may be written as a combination of products of $N_c$ mesons $M$,
but $B$ is an independent operator, charged under $\Z_2^{\cal C}$.
For $G=O(N_c)$ and $G=Pin(N_c)$ the operator $B$ does not exist, since the charge conjugation symmetry is gauged.

Consider now a monopole operator like $Y$ in $SO(N_c)$, which breaks
\eqn\ybreaking{SO(N_c)\to S\Big(O(N_c-2)\times O(2)\Big).}
Note that this includes transformations with determinant $(-1)$ both in
$SO(N_c-2)$ and in $SO(2)$. This means that the gauge-invariant operator $Y$ must be charge conjugation even in $SO(2)$, and thus it reduces to the operator $W_+$ \wpmdef\ in this group. However, we can also build a ``baryon-monopole'' operator involving $W_-$ in $SO(2)$ \wpmdef, by defining
\eqn\betadef{\beta = Q^{N_c-2} W_-,}
with the indices of the flavors contracted by an epsilon symbol in the $SO(N_c-2)$ that is unbroken by the monopole. The product $Q^{N_c-2}$ is invariant under the $SO(N_c-2)\times SO(2)$ subgroup of $SO(N_c)$ that is left unbroken by the monopole, and its product with $W_-$ is invariant also under the extra $\Z_2$, so that \betadef\ defines a gauge-invariant operator in $SO(N_c)$.
Note that the operators \betadef\ exist for any $N_f \geq N_c-2$. The standard matching of quantum numbers for monopole operators, generalizing \vevrel, implies that when we give a VEV to $Q$ breaking $SO(N_c)$ to $SO(2)$, $\beta$ reduces directly to $W_{-low}$ in the low-energy $SO(2)$ theory (with no extra factors of $\vev{Q}$).

As we discussed above, when the gauge group is $Spin(N_c)$, the monopole operators $Y$ and $W_\pm$
do not exist. But we can still repeat the above discussion using the operator $Y_{Spin}$ (which reduces to ${\hat W}_+$), and define a baryon-monopole
\eqn\betaspindef{\beta_{Spin} = Q^{N_c-2} {\hat W}_-}
as above. Note that in an $SO(2)$ theory, ${\hat W}_- = V_+^2 - V_-^2 = (V_+ - V_-) (V_+ + V_-)$, so in $SO(N_c)$ theories $\beta_{Spin} = \beta Y$.

The operator $\beta$ satisfies an interesting chiral ring relation.  Consider the $3d$ $SO(N_c)$ gauge theory with $N_f=N_c-2$. At a generic point on the moduli space of this theory, we break $SO(N_c) \to SO(2)$. The monopole operator $Y$ reduces in the low-energy $SO(2)$ theory to $W_{+low}$; the standard mapping of monopoles \vevrel\ implies that $W_{+low} = Y \sqrt{\det(M)}$. The low-energy $SO(2)$ theory has no massless flavors, and hence $V_{+low} V_{-low} = 1$.  Therefore, in this vacuum
\eqn\betarela{\beta^2 = W_{-low}^2 = V_{+low}^2 - 2 V_{+low} V_{-low} + V_{-low}^2 = W_{+low}^2 - 4 = Y^2 \det(M) - 4.}
This reflects an exact chiral ring relation
\eqn\betarel{\beta^2 = Y^2 \det(M) - 4 ~,}
which is valid in every vacuum of this theory.
Classically this theory has a point at the origin of its moduli space where $Y=M=\beta=0$, but we see that quantum mechanically the moduli space is deformed and obeys \betarel.  This is similar to the deformation of the classical moduli space in some $4d$ theories \SeibergBZ\ and in some $3d$ theories \AharonyBX.

Similarly, we can use the extra monopole operators of $SO(4)$, by having monopoles breaking
$SO(N_c)$ to $SO(N_c-4)\times U(2)$. More precisely, to define such monopoles we first choose some $S(O(N_c-4)\times O(4))\subset SO(N_c)$, and then turn on a monopole like \zdef\ in the $SO(4)$ factor, breaking it to $U(2)$. As in our discussion above, the monopole $Z$ on its own reduces to the even operator $Y_+$ in $SO(4)$ \ypmdef. But we can now consider instead the operator
\eqn\bdef{b = Q^{N_c-4} Y_-,}
where the quarks are contracted with the epsilon symbol of $SO(N_c-4)$. As in our discussion above, this operator $b$ is gauge-invariant in the $G=SO(N_c)$ theory, including also the gauge transformations of determinant $(-1)$ in the two factors. The same discussion applies also to $Spin(N_c)$ theories. The baryon-monopole $b$ exists for any $N_f \geq N_c-4$ (for $N_c=4$ it is simply $Y_-$). As in our discussion of $Z$ above, due to quantum effects $b$ is not really a chiral operator in $3d$ SQCD theories, but we will see that it still plays a role in our analysis.

We cannot generalize this construction to breaking $SO(N_c)\to S\Big(O(N_c-2n)\times O(2n)\Big)$ with $n>2$, because there is no obvious monopole operator in $SO(2n)$ that is odd under charge conjugation.

\subsec{On $3d$ $O(N_c)$ therories}

We mentioned above that one can obtain $O(N_c)$ theories by gauging the charge conjugation symmetry of $SO(N_c)$, but in fact there are two different
$O(N_c)$ theories in $3d$ that will play a role in this paper. In one $O(N_c)$ theory the minimal monopole operator $Y$ of $SO(N_c)$, which is charge-conjugation-even in the $SO(N_c)$ theory, is
gauge-invariant, and is the minimal monopole operator also for $O(N_c)$. We will denote this theory by $O(N_c)_+$; this is the theory that was discussed
in previous papers about $3d$ $O(N_c)$ theories and their dualities. However, as in \AST, one can also define a second $O(N_c)_-$ theory, in which the
monopole operator $Y$ is charge-conjugation-odd (it changes sign under gauge transformations whose determinant is $(-1)$). In this $O(N_c)_-$ theory,
$Y$ and $B$ are both not gauge-invariant, but their product, as well as the operators $Y_{Spin}$ and $\beta$, are gauge-invariant (note that $\beta$
is not gauge-invariant in the standard $O(N_c)_+$ theory).

In the Lagrangian language, the two theories differ by a discrete theta angle, analogous to the one that distinguishes the $4d$ $SO(N_c)_\pm$ theories \AST.
The relevant term in the Lagrangian is proportional to $w_1\wedge w_2$, where $w_i \in H^i(X,\Z_2)$ are $\Z_2$-valued characteristic classes of the $O(N_c)$ bundle on a manifold $X$; 
$w_1$ is non-zero when the $O(N_c)$ bundle cannot be written as an $SO(N_c)$ bundle, while $w_2$ is non-zero when
the $O(N_c)$ bundle cannot be written as a $Pin(N_c)$ bundle. In particular, $w_2$ is non-zero on a two-sphere around an insertion of the 
operator $Y$. Note that the two options exist only for $O(N_c)$ gauge groups, not for $SO(N_c)$, $Spin(N_c)$ or $Pin(N_c)$ (in which either $w_1$, or
$w_2$, or both, are trivial).\foot{More generally, it may be possible to add other terms such as $w_1\wedge w_1\wedge w_1$, but we will not discuss this here.}

\newsec{The Coulomb branch of $4d$ $so(N_c)$ theories on $\S^1$}

\subsec{$4d$ $Spin(N_c)$ theories on $\S^1$}

As discussed in \ARSW, when one compactifies a $4d$ gauge theory on $\S^1$ and goes to low
energies, naively one gets the same gauge theory in $3d$, but there are two important differences.
The first is that the Coulomb branch coordinates now come from holonomies of the gauge field on
a circle, so the coordinates $\sigma$ described above are periodic and the Coulomb branch is compact.
The second is that there is an extra monopole-instanton in the theory on a circle, that gives an
extra term in the effective superpotential. We will start by discussing these aspects for the reduction of a $4d$ theory with $G=Spin(N_c)$, and consider $G=SO(N_c)$ theories in the next subsection. We begin with the case of $N_c \geq 5$.

In the $4d$ theory on a circle of radius $r$, the scalars $\sigma$ described in the previous section originate from $A_3$, but only the eigenvalues of $U = P \exp(i\oint A_3)$ are gauge-invariant, so there are relations between different values of $\sigma$ associated with large gauge transformations. In particular, the eigenvalues of $U$ in the vector representation are $\exp(\pm 2\pi i r \sigma_i)$, so each $\sigma_i$ gives the same holonomy in this representation as $\sigma_i+1/r$. Note that the
eigenvalues of $U$ in the spinor representation are $\exp(2\pi i r (\pm \sigma_1/2 + \cdots))$, so the periodicity of each $\sigma_i$ for $G=Spin(N_c)$ is actually $2/r$ (or one can shift two $\sigma_i$'s together by $1/r$). But the masses of W-bosons and matter fields in the vector representation are periodic in $\sigma_i$ with periodicity $1/r$. In particular, whenever all the $\sigma_i$ are integer multiples of $1/r$, the gauge group is unbroken and any matter fields in the vector representation are massless.

In the $4d$ theory on a circle, we can get an enhanced non-Abelian symmetry not just by having
$\sigma_i \to \sigma_{i+1}$, but also by having eigenvalues meet the images of other eigenvalues. When $\sigma_1$ meets the image $-\sigma_1$ at $\vev{\sigma_1}=1/2r$ there is no enhanced non-Abelian symmetry, since we just have an enhancement of $U(1)$ to $SO(2)$ or $Spin(2)$. However, when $\sigma_1$ meets the image
of $-\sigma_2$, when $\sigma_1=-\sigma_2+1/r$, there is an enhancement of $U(1)^2$ to $U(2)$ (if this happens at $\sigma_1=1/2r$ then there is even an enhancement to $SO(4)$ or $Spin(4)$). The same computation yielding the monopole-instanton contributions described above \AffleckAS, thus gives in the theory on a circle an extra superpotential. The analogy with the $1/Y$ superpotential of \AffleckAS\ implies that semi-classically the extra superpotential looks like
\eqn\newsuppot{{1\over {\exp\left({{{1\over r}-\sigma_1-\sigma_2}\over{{\hat g}_3^2}}-i(a_1+a_2)\right)}} \approx \eta Z,}
where $Z$ was defined in \zdef, and $\eta \equiv \Lambda^{b_0} = \exp(-8\pi^2/ g_4^2) = \exp(-4\pi / g_3^2 r)$ is the strong coupling scale of the $4d$ gauge theory ($b_0=3(N_c-2)-N_f$ is the $4d$ one-loop beta function coefficient, and we set the $4d$ theta angle to zero and the renormalization scale to one for simplicity). The precise form \newsuppot\ follows by carefully analyzing all the instantons, as in \refs{\SeibergNZ\LeeVP,\LeeVU,\DaviesUW,\DaviesNW}. From the point of view of the effective $3d$ theory, \newsuppot\ breaks precisely the global $U(1)$ symmetry that is anomalous in the $4d$ theory. Note that in the $3d$ theory $Z$ is not a chiral operator, but in the $4d$ theory on a circle, it can no longer be written as a descendant, due to the extra superpotential \newsuppot.

In the pure SYM theory, the extra term \newsuppot\ stabilizes the runaway caused by \quantumeven\ and \quantumodd, and leads to a finite number of supersymmetric vacua, obtained by solving the F-term equations for the $Y_i$. One can check  that, both for even and odd values of $N_c$, this leads to $(N_c-2)$ supersymmetric vacua, with
\eqn\zvev{(2 \eta \vev{Z})^{N_c-2} = 16 \eta.}
This is the same number of vacua
as in the $4d$ theory, as expected in this case \AST, and the value of the superpotential also agrees with its $4d$ value. As discussed in \ARSW, the $4d$ chiral operator $S\propto {\rm tr}(W_{\alpha}^2)$ reduces in the theory on a circle to $Z$, with a chiral ring relation
\eqn\szrelation{S = \eta Z,}
which is consistent with \zvev. Note that in the $4d$ theory on a circle, the monopole and baryon-monopole operators discussed in the previous section do not exist microscopically (due to the compactness of the Coulomb branch), but we can give a microscopic definition to $Z$ using \szrelation.

Moving next to the theories with flavors,
note that unlike in the cases of $G=SU(N_c)$ and $G=USp(2N_c)$ discussed in \ARSW, here the extra superpotential \newsuppot\ is not proportional to the coordinate $Y_{Spin}$ along the unlifted Coulomb branch of the $3d$ $Spin(N_c)$ SQCD theory. Thus, this superpotential does not lift the Coulomb branch, but just adds extra interactions. The coordinate $Y_{Spin}$ on this Coulomb branch is uncharged under the continuous $SU(N_f)\times U(1)_R$ global symmetry that is preserved by \newsuppot.

The global structure of the moduli space is interesting. First, $\sigma_1$ should be identified with $(-\sigma_1+2/r)$ because they lead to the same holonomy, so we can take $0 \leq \sigma_1 \leq 1/r$. In the quantum theory the moduli space is parameterized by $Y_{Spin}$ \ydef, and we identify the point $Y_{Spin}=0$ with $\sigma_1=0$, and the point $Y_{Spin}=\infty$ with $\sigma_1=1/r$.  Classically, at generic values of $Y_{Spin}$ the $Spin(N_c)$ symmetry is broken, and at the two special points $Y_{Spin}=0,\infty$ the full $Spin(N_c)$ symmetry is preserved.

Second, if our $Spin(N_c)$ gauge theory does not couple to matter fields in a spinor representation, the compactified theory has a {\it global $\Z_2$ symmetry}, acting on the moduli space by $\sigma_1\to -\sigma_1 + 1/r$. The point is that these two values of $\sigma_1$ represent two different holonomies in $Spin(N_c)$, but this difference is not felt by any dynamical field.\foot{More generally, whenever we have a $\Z_p$ gauge theory in $d$ dimensions that does not couple to charged fields, the theory compactified to $d-1$ dimensions has both a gauge $\Z_p$ symmetry and a global $\Z_p$ symmetry.  One way to see that is to represent the $\Z_p$ gauge theory by a one-form $U(1)$ gauge field $A^{(1)}$ and a $(d-2)$-form gauge field $A^{(d-2)}$ with a Lagrangian given by ${p\over 2\pi} A^{(1)} \wedge dA^{(d-2)}$ \refs{\MaldacenaSS,\BanksZN}.  Reducing this system to $d-1$ dimensions leads to four fields, with the Lagrangian ${p\over 2\pi}\left( A^{(1)} \wedge dA^{(d-3)}+A^{(0)} \wedge dA^{(d-2)}\right)$.  The first term represents a $\Z_p$ gauge theory, and the second term describes a $\Z_p$ global symmetry (and a gauge symmetry for $A^{(d-2)}$). The global symmetry acts as $A^{(0)} \to A^{(0)} + 2\pi/p$, with the identification $A^{(0)} \sim A^{(0)} + 2\pi$.  If the original $d$ dimensional theory includes matter fields charged under the $\Z_p$ gauge symmetry, the global symmetry $A^{(0)} \to A^{(0)} + 2\pi/p$ in the $d-1$ dimensional theory is explicitly broken, but the identification $A^{(0)} \sim A^{(0)} + 2\pi$ is preserved.  A famous example of this phenomenon, due to Polyakov, is the compactification of a $4d$ $SU(N)$ gauge theory without matter on $\S^1$.  The resulting $3d$ theory has a global $\Z_N$ symmetry, which originates from the $4d$ $\Z_N\subset SU(N)$ gauge symmetry.  The order parameter for its spontaneous breaking is the Polyakov loop $e^{i A^{(0)}} \propto {1\over N} {\rm tr}(e^{i \oint A})$. In our case, the relevant $\Z_2$ gauge symmetry is the subgroup of the center of $Spin(N_c)$ that acts on spinors.}
The two special points on the moduli space $Y_{Spin}=0,\infty $ are not identified.  Instead, they are exchanged by the global $\Z_2$ symmetry, which acts on the moduli space as
\eqn\globaltrans{Y_{Spin} \longleftrightarrow {1\over {\eta^2 Y_{Spin}}}.}

In the $4d$ $G=Spin(N_c)$ theory there are several baryonic operators defined in \refs{\SeibergPQ,\IntriligatorID},
\eqn\baryonsfourd{B = Q^{N_c},\qquad {\cal W}_{\alpha} = W_{\alpha} Q^{N_c-2}, \qquad b_{4d} = W_{\alpha}^2 Q^{N_c-4},}
which all involve contractions with the epsilon symbol of $Spin(N_c)$. The first operator obviously
reduces to the same baryon operator in $3d$. The second operator is useful when $(N_c-2)$ quarks
get expectation values breaking the gauge group to $g=so(2)$, to label the remaining unbroken $so(2) \subset so(N_c)$; in the effective $3d$ theory the same role is played by $\beta$ or $\beta_{Spin}$. Similarly, when the gauge group is broken to $g=so(4)$, we have a relation between $S_1-S_2$ (of $SU(2)\times SU(2)$ in $4d$) and $Y_-$ of $so(4)$ \ypmdef\ that is analogous to \szrelation, and that implies that the $4d$ baryon $b_{4d}$ goes down in the low-energy effective theory to $\eta$ times the baryon-monopole $b$ defined in \bdef.

As in the $3d$ discussion, there are some modifications to this story for low values of $N_c$. For $N_c=2$ there is no extra superpotential on the circle, and the moduli space for $G=U(1)$ was discussed in section 4.2 of \ARSW. For $N_c=3$, there is only a single non-trivial eigenvalue of $\sigma$, and instead of \newsuppot\ we get a superpotential from the fact that when $\sigma_1 \to 1/r$ the gauge group is enhanced again to $so(3)$. The full superpotential of the $4d$ $Spin(3)$ pure SYM theory on a circle takes the form $W = 1 / Y_{Spin} + \eta^2 Y_{Spin}$, consistent with the global symmetry \globaltrans\ that acts also in this case.
Note that in this special case the instanton factor of $Spin(3)$ is actually $\eta_{Spin(3)} = \eta^2 = \Lambda^{2b_0}$ (keeping our general definition above for $\eta$), so this discussion is consistent with the standard discussion of $4d$ $SU(2)$ theories on a circle \SeibergNZ. The $4d$ $Spin(3)$ SYM theory on a circle has two vacua
at $Y_{Spin} = \pm 1 / \eta = \pm 1 / \sqrt{\eta_{Spin(3)}}\ ,$  which are fixed points of the $\Z_2$ symmetry \globaltrans\ (though they are related by a $\Z_4$ global R-symmetry transformation).

For $N_c=4$ with $Spin(4)=SU(2)\times SU(2)$, we have in the pure $4d$ gauge theory on a circle two copies of the discussion of the previous paragraph,
\eqn\sofourw{W = {1\over Y^{(1)}} + {1\over Y^{(2)}} + \eta Y^{(1)} + \eta Y^{(2)}}
(consistent with the identification of $Z$ discussed in section \baryonmon). Note that the two $SU(2)$'s have $\eta_1 = \eta_2 = \eta$. There are $4$ supersymmetric  vacua at $Y^{(1)} = \pm 1 / \sqrt{\eta}$, $Y^{(2)} = \pm 1 / \sqrt{\eta}$. In two of these vacua
$Y_{Spin} = Y^{(1)} Y^{(2)} = 1 / \eta$, and in the other two $Y_{Spin} = -1 / \eta$; they are all
fixed points of the global symmetry transformation \globaltrans.

\subsec{$4d$ $SO(N_c)$ theories on $\S^1$}

We saw in the previous section that an important difference between $G=Spin(N_c)$ and $G=SO(N_c)$
is that in the latter case there is an extra monopole operator $Y$ that can be used to label the
Coulomb branch. The Coulomb branch of the $SO(N_c)$ theory is a double cover of that of the $Spin(N_c)$ theory, which is labeled by $Y_{Spin} = Y^2$. When we discuss the $4d$ theory on a circle, another difference is that the global symmetry transformation \globaltrans\ becomes a large gauge transformation in $SO(N_c)$, with the gauge transformation parameter periodic around the circle in $SO(N_c)$ but not in $Spin(N_c)$. Thus, the points on the Coulomb branch related by \globaltrans\ are identified. This is related to the fact that we no longer have a spinor Wilson line, so the $\sigma_i$ have periodicity $1/r$, and we can restrict the range of $\sigma_1$ to $0 \leq \sigma_1 \leq 1/2r$.

Let us see in more detail how the $\Z_2$ large gauge transformation acts on $Y$. Given the relation $Y^2 = Y_{Spin}$, and the action on $Y_{Spin}$ \globaltrans, there are two possible actions on $Y$: it could act either as $Y \longleftrightarrow 1 / \eta Y$,
or as $Y \longleftrightarrow -1/\eta Y$.  These two options are related to the fact that there are two distinct $SO(N_c)$ $4d$ gauge theories \AST, $SO(N_c)_\pm$. As explained in \AST\ and reviewed in the introduction, the difference between the three $4d$ gauge theories $Spin(N_c)$ and $SO(N_c)_\pm$ is in the choice of line operators.  $Y_{Spin}$ has the same magnetic quantum numbers as the minimal $4d$ 't Hooft loop of the $Spin(N_c)$ theory $H^2$, wrapped on the circle.  $Y$ in the $SO(N_c)_+$ theory is related to the wrapped $4d$ 't~Hooft loop $H$, and $Y$ in the $SO(N_c)_-$ theory is related to the wrapped $4d$ 't Hooft-Wilson loop $HW$. (The wrapped 't Hooft loops are not BPS operators, but they reduce to chiral superfields in the low-energy effective action.) Correspondingly, the large gauge transformation in $SO(N_c)_\pm$  acts on $Y$ and leads to the identification (recalling that the wrapped spinor Wilson line $W$ is odd under this transformation)
\eqn\songauge{SO(N_c)_+ \ : \qquad Y \sim {1 \over {\eta Y}}~,\qquad\qquad
SO(N_c)_- \ : \qquad Y \sim {-{1 \over {\eta Y}}}~ .}
In the $3d$ dimensionally reduced theory which arises as $\eta \to 0$, we are left just with the region near $Y=0$ so there is
no longer any distinction between $SO(N_c)_+$ and $SO(N_c)_-$, and there is just a single $3d$
$SO(N_c)$ gauge theory.

In both cases, in the theories with flavors, the Coulomb branch of the $4d$ gauge theory on a circle is labeled by $Y$ with the identification \songauge. These theories also have a $\Z_2$ global symmetry taking
$Y \to -Y$, which acts on the Coulomb branch (this symmetry, acting on the non-trivial wrapped 't~Hooft lines, is not present in the $Spin(N_c)$ theory). Note that this symmetry remains also in the $3d$ limit, and that the baryon-monopole $\beta$ of \betadef\ is odd under it (while $\beta_{Spin}$ of \betaspindef\ is even).

To summarize, in the $SO(N_c)_\pm$ theories we have a gauge identification on the Coulomb branch given by \songauge, and in all $3$ theories we have a
$\Z_2$ global symmetry changing the sign of the non-trivial line operator wrapped on the circle. We denote this symmetry by $\Z_2^{\cal M}$. It acts on the Coulomb branch as:
\eqn\globalsymm{Spin(N_c)\ : \qquad Y_{Spin} \longleftrightarrow {1\over {\eta^2 Y_{Spin}}}~,\qquad\qquad SO(N_c)_{\pm}\ : \qquad Y \longleftrightarrow -Y~.}

We can now look at the pure $so(N_c)$ SYM theory on a circle, and see how many vacua we have
in the different $so(N_c)$ theories \AST. Solving the F-term equations of the $Spin(N_c)$ theory as
above gives $(N_c-2)$ different solutions for $Z$ \zvev, which all have $\vev{Y_{Spin}} = \vev{Y_1 Z} = 1 / \eta$.
In the $Spin(N_c)$ theory, these are all fixed points of $\Z_2^{\cal M}$. In the $SO(N_c)$
theories, we can additionally characterize each vacuum by the expectation value of $Y$, which can
be $\pm 1 / \sqrt{\eta}$. In the $SO(N_c)_+$ theory, each of the two possibilities maps to itself under the
gauge identification \songauge, and thus each solution of \zvev\ splits into two separate
vacua, giving a total of $2(N_c-2)$ supersymmetric vacua. On the other hand, in the $SO(N_c)_-$
theory, the two options for the sign of $Y$ are mapped to each other by the gauge identification \songauge,
and thus there are only $(N_c-2)$ supersymmetric vacua. This shows the consistency of our identifications
\songauge\ with the counting of vacua presented in \AST.

While much of our discussion remains the same also for low values of $N_c$, it is interesting to verify its consistency with the relations \sothree\ and \sofour. For $N_c=3$ we mentioned that $\eta_{SO(3)} = \eta^2$, so shifting the $SO(3)$ theta angle by $2\pi$ as in \sothree\ takes $\eta \to -\eta$,  consistent with \songauge. In this case there is no extra monopole operator $Z$, and $Y$ carries the full information about the vacuum. Each of the $Spin(3)$ vacua at $Y_{Spin} = \pm 1 / \sqrt{\eta_{Spin(3)}}$ splits into two possible $SO(3)$ vacua at
$Y = \pm \sqrt{Y_{Spin}}$. In the $4d$ $SO(3)_+$ SYM theory on a circle two of these are identified by the large gauge transformation \songauge, and the other two are not, and the situation is reversed in the $SO(3)_-$ theory. Both theories thus have $3$ vacua \AST. Similarly, for $N_c=4$ the two vacua at $Y_{Spin}=1/\eta$ each split into two vacua with $Y = \pm 1 / \sqrt{\eta}$, and the two vacua at $Y_{Spin} = -1/\eta$ each split into two vacua with $Y = \pm i / \sqrt{\eta}$. In the $SO(4)_+$ theory the latter two vacua are identified by \songauge, so that we have $6$ overall vacua, and similarly in the $SO(4)_-$ theory the vacua from the first splitting are identified, and again we have $6$ overall vacua \AST. The transformation shifting (say) the first theta angle in \sofour\ changes the sign of $Y^{(1)}$ and thus of $Y_{Spin}$, and exchanges the two types of vacua, consistent with its exchanging $SO(4)_+$ and $SO(4)_-$.

\newsec{The $4d$ duality on a circle}

\subsec{Dual theories on a circle}

As discussed in \ARSW, whenever we take two theories that are IR-dual in $4d$, compactify them on a circle of radius $r$, and go to low energies (compared to the scales $1/r$, $\Lambda$ and ${\tilde \Lambda}$), the resulting low-energy $3d$ theories are IR-dual as well. We can start from the $4d$ duality reviewed in the introduction (with $N_c \geq 4$ and $N_f \geq N_c$), between theory A with gauge group $G$ (which can be $Spin(N_c)$, $SO(N_c)_+$ or $SO(N_c)_-$), with $N_f$ chiral multiplets $Q^i$ in the vector representation and $W_A=0$, and theory B with gauge group ${\tilde G}$ (which is $SO({\tilde N}_c)_-$, $SO({\tilde N}_c)_+$ or $Spin({\tilde N}_c)$, respectively, with ${\tilde N}_c=N_f-N_c+4$), with $N_f$ chiral multiplets $q_i$ in the vector representation, $N_f(N_f+1)/2$ singlets $M^{ij}$ and a superpotential $W_B = {1\over 2} M^{ij} q_i q_j$.\foot{We begin by discussing connected gauge groups, we will discuss the disconnected $O(N)$ and $Pin(N)$ cases below.} The continuous global symmetries in both theories are $SU(N_f)\times U(1)_R$, with $Q$ in the $({\bf N_f})_{((N_f-N_c+2)/N_f)}$ representation in theory A, and $q$ in the $({\bf {\overline N_f}})_{((N_c-2)/N_f)}$ representation in theory B. Both theories have a
$\Z_{2N_f}\times \Z_2^{\CC}$ discrete symmetry. The charge conjugation symmetry $\Z_2^\CC$, generated by $\CC$, under which the baryonic operators \baryonsfourd\ are odd, maps to itself under the duality. The $\Z_{2N_f}$ symmetry of theory A (which is a subgroup of the anomalous axial $U(1)$ symmetry) is generated by $g : Q \to \exp(2\pi i / 2 N_f) Q$, and that of theory B by ${\tilde g} : q\to \exp(-2\pi i / 2 N_f) q$. They are mapped by the duality as $g \leftrightarrow {\tilde g} \CC$ \refs{\SeibergPQ,\IntriligatorID}.

When we compactify the two theories on a circle, we generate extra superpotentials
\eqn\circlesuppot{W_A = \eta Z,\qquad\qquad W_B = {1\over 2} M^{ij} q_i q_j + {\tilde \eta} {\tilde Z},}
where the $4d$ duality implies $\eta {\tilde \eta} = (-1)^{N_f-N_c} / 256$ \refs{\IntriligatorID,\IntriligatorAU}. We also get the compact Coulomb branches on both sides, described in the previous section. The general arguments of \ARSW\ imply that the low-energy theories with these extra superpotentials and compact Coulomb branches should be dual at low energies.

The mapping between the chiral operators involving the flavors, and the associated flat directions, is the same in the theory on a circle as in $4d$; $M^{ij}$ in theory B is identified with $Q^i Q^j$, and the baryonic operators $B$, ${\cal W}_{\alpha}$ and $b_{4d}$ are identified with ${\tilde b}_{4d}$, ${\tilde {\cal W}}_{\alpha}$ and $\tilde B$, respectively \refs{\SeibergPQ,\IntriligatorID}. But on a circle we have the extra Coulomb branch that we need to identify, which is not lifted by the extra superpotential \circlesuppot, and we have the associated chiral operators that parameterize it. Note that the Coulomb branch coordinates are not charged under any continuous global symmetries, so these do not constrain their mapping.

However, there are discrete symmetries acting on the Coulomb branch \globalsymm, and these should map to each other under the duality (this follows from the mapping of the corresponding $4d$ non-trivial line operators \refs{\StrasslerFE,\AST}). In fact, the full mapping of the Coulomb branches is uniquely determined by requiring that we have a single-valued meromorphic transformation between them (after identifying by the large gauge transformations \songauge\ for $SO(N)$), which correctly maps the global $\Z_2$ symmetries \globalsymm, together with the extra requirement (for the $SO(N)_+$ case) that the mapping is non-trivial (this follows from the analysis of the next subsection). The mapping between the Coulomb branch coordinate $Y_{Spin}$ of $Spin(N_c)$, and the coordinate ${\tilde Y}$ of the dual $SO({\tilde N}_c)_-$ theory, takes the form
\eqn\firstmap{\eta Y_{Spin} = \left[ {{i + \sqrt{\tilde \eta} {\tilde Y}}\over {1 + i \sqrt{\tilde \eta} {\tilde Y}}} \right]^2.}
The two choices of sign for $\sqrt{\tilde \eta}$ are related by the global symmetry ${\tilde Y} \to -{\tilde Y}$, which maps in the dual theory to the global symmetry $Y_{Spin} \to 1 / \eta^2 Y_{Spin}$. The two identified points ${\tilde Y}$ and $(-1 / {\tilde \eta} {\tilde Y})$ map to the same value of $Y_{Spin}$, as they should.

The map between the coordinate $Y$ of the $SO(N_c)_-$ theory, and the coordinate ${\tilde Y}_{Spin}$ of the dual $Spin({\tilde N}_c)$ theory, is simply the inverse of this map (consistent with the fact that performing the duality twice should bring us back to the same point). The map between the two $SO(N)_+$ theories is just the square root of \firstmap,
\eqn\secondmap{{\sqrt \eta} Y = {{i + \sqrt{\tilde \eta} {\tilde Y}}\over {1 + i \sqrt{\tilde \eta} {\tilde Y}}}.}
Again, one can check that it is consistent with the discrete gauge and global symmetries on both sides.

For low numbers of flavors, $N_c=3$ or ${\tilde N}_c=3$, there are extra terms appearing in the dual superpotential \IntriligatorID, and the mapping of $\eta$ to ${\tilde \eta}$ is somewhat different because of the different instanton factor in $so(3)$ theories. Both for $N_c=3$ and $N_c=4$ there are also extra ``triality'' relations between the $4d$ theories \refs{\IntriligatorID,\IntriligatorER,\AST}, because of the relations \sothree\ and \sofour. In any case, the extra superpotential factors and dualities do not introduce any new issues when reduced on a circle, so we will not discuss them further here.

Note that we can get the dualities for the $O(N)$ and $Pin(N)$ theories just by gauging the charge conjugation symmetries in the dualities for $SO(N)$ and $Spin(N)$. This is true both in $4d$, and for the $4d$ theories on a circle. The main difference in the non-connected cases is that we do not have the baryon operators on both sides, so we have fewer distinguishable vacua and fewer chiral operators.

\subsec{A test of the duality and of the Coulomb branch mapping}

As a consistency check for our mappings \firstmap, \secondmap, let us analyze what happens far on the Higgs branch of theory A. In this theory we can turn on a vacuum expectation value (VEV) for $M^{ij}=Q^i Q^j$ of rank $N_c$, such that the gauge group is completely broken. For each such VEV there are two supersymmetric vacua, differing by the sign of the non-zero component of the baryon $B=Q^{N_c}$ (which squares to $\det_{N_c\times N_c}(M)$). This is true both in $4d$ and in the theory on a circle.  In the latter case, since the gauge group is completely broken by the quarks, we must be at a point on the Coulomb branch where classically the gauge group is unbroken, namely $Y_{Spin}=0$ or $Y_{Spin}=\infty$ for $Spin(N_c)$, and $Y=0$ for $SO(N_c)$.

 We now discuss these vacua in theory B.  The meson VEV gives a mass to $N_c$ quarks, leaving $(N_f-N_c)={\tilde N}_c-4$ massless quarks.  The low-energy theory is $so({\tilde N_c})$ with ${\tilde N_c}-4$ massless flavors $q$, with a scale ${\tilde \eta}_{low} = {\tilde \eta} \, \det_{N_c\times N_c}(M)$. Let us first ignore the singlets and the superpotential. We then have classically a moduli space for the $q$'s, where generically the gauge group is broken to $so(4)$, with no light charged fields. We can think of the $so(4)$ theory as an $su(2)\oplus su(2)$ gauge theory, where each $su(2)$ factor has an instanton factor ${\tilde \eta}_{su(2)}$ related to that of the original $so({\tilde N_c})$ theory by ${\tilde \eta}_{su(2)} = {\tilde \eta}_{low} / \det(q q)$. In the $4d$ theory, gaugino condensation in each $su(2)$ factor leads to an effective superpotential $W=2(\pm 1 \pm 1) {\tilde \eta}_{su(2)}^{1/2}$ for the $q$'s, which vanishes (and leads to a supersymmetric vacuum for the original $so({\tilde N}_c)$ theory), if and only if we choose opposite signs for the two gaugino condensates. The operator ${\tilde b}_{4d}$ of theory B is then equal to
\eqn\bvev{{\tilde b}_{4d} = q^{{\tilde N_c}-4} {\tilde W}_{\alpha}^2 = \pm 2 \sqrt{\det(q q)} \sqrt{{\tilde \eta}_{su(2)}},}
where at generic points on the moduli space this ${\tilde W}_\alpha^2$ is the difference between the two gaugino bilinears of the two $su(2)$ factors.  It does not vanish in the SUSY vacua. This operator obeys
\eqn\bsquared{{\tilde b}_{4d}^2 = 4 \, \det(q q) \, {\tilde \eta}_{su(2)} = 4 {\tilde \eta}_{low} = 4 {\tilde \eta} \, \det_{N_c\times N_c}(M).}
Clearly, this relation is true for any value of the $q$'s and is an exact ring relation.  Hence, in the full $4d$ theory B we find that ${\tilde b}_{4d}$ is non-zero when $M$ has rank $N_c$, and obeys a similar relation to $B$ of theory A, so that we can identify
\eqn\bBrel{{\tilde b}_{4d} = 2 \sqrt{\tilde \eta} B}
(when we normalize the superpotential of theory B to be ${1\over 2} M q q$ with no extra factors). We thus identify the two supersymmetric vacua discussed above for this VEV of $M$ in theory B. Similarly, one can show that if ${\rm rank}(M) > N_c$ there is no supersymmetric vacuum in theory B.

Let us now repeat our discussion of theory B, when it is compactified on a circle.  Most of the discussion is the same for $Spin(N_c)$ and $SO(N_c)$, so first we will not distinguish between them.
Again we turn on a VEV of rank $N_c$ for $M$, leaving in theory B $({\tilde N_c}-4)$ massless flavors. The matching between the high-energy and low-energy Coulomb branch coordinates implies that ${\tilde Z}_{low} = {\tilde Z} / \det_{N_c\times N_c}(M)$ (as in \massrel), so that the low-energy superpotential \circlesuppot\ includes ${\tilde \eta} {\tilde Z} = {\tilde \eta}_{low} {\tilde Z}_{low}$. Again, let us ignore for a moment the extra ${1\over 2} M q q$ superpotential in theory B, and imagine that we turn on generic VEVs for the remaining massless $q$'s, breaking the gauge symmetry to $so(4)$. Each of the $su(2)$ factors in $so(4)$ now has a Coulomb branch coordinate ${\tilde Y}^{(j)}$, and, as in \vevrel, the relation of the low-energy and high-energy coordinates is
\eqn\yvevrels{{\tilde Y}_+ = {\tilde Y}^{(1)}+{\tilde Y}^{(2)} =
{\tilde Z}_{low} \, \det(q q).}
The full low-energy superpotential, including the Affleck-Harvey-Witten superpotentials \AffleckAS\ of the two $su(2)$'s, is thus
\eqn\fullsofour{W_B = \frac{1}{{\tilde Y}^{(1)}} + \frac{1}{{\tilde Y}^{(2)}} + \frac{{\tilde \eta}_{low}}{\det(q q)} ({\tilde Y}^{(1)} + {\tilde Y}^{(2)}),}
leading to four states with
\eqn\tildeyvev{{\tilde Y}^{(j)} = \pm \sqrt{{\det(q q) \over {\tilde \eta}_{low}}} = \pm {1 \over \sqrt{{\tilde \eta}_{su(2)}}}~.}
(More precisely, in counting the physical states we should take into account the global aspects of whether our gauge group is $Spin(\tilde N_c)$ or $SO(\tilde N_c)$.  We will do this momentarily.) Note that this is consistent with our discussion in the previous paragraph, and with the relation $\eta Y = S$ for $SU(2)$ theories on a circle \ARSW; in this case we have (in the chiral ring) ${\tilde \eta}_{su(2)} {\tilde Y}^{(j)} = S_j$. As in $4d$, in order not to turn on a superpotential for the $q$'s we need the expectation values to obey ${\tilde Y}^{(1)} = - {\tilde Y}^{(2)}$.
So, as in $4d$, we find two supersymmetric vacua in theory B, in which
\eqn\tiYone{{\tilde Y}^{(1)} {\tilde Y}^{(2)} = - {1 \over {\tilde \eta}_{su(2)}} = - {\det(q q)\over {\tilde \eta}_{low}}~.}

Note that the two choices for the sign of ${\tilde Y}^{(j)}$ are distinguished by the sign of the baryons ${\tilde b} = q^{{\tilde N}_c-4} {\tilde Y}_-$.\foot{Like the operator $Z$, the $b$ baryon-monopole operators in the low-energy effective action cannot be written as descendants in the $4d$ theory on a circle, even though they are descendants in $3d$ SQCD.} The discussion above implies that ${\tilde b}^2$ has an expectation value equal to $4 \, \det_{N_c\times N_c}(M) / {\tilde \eta}$, so we can identify
\eqn\tildebBrel{{\tilde b}={1\over {\tilde \eta}}{\tilde b}_{4d} ={2\over \sqrt{\tilde \eta} } B ~.}
For every value of $M$ of rank $N_c$ in theory A there are two choices of $B$, and they are mapped under the duality to the two choices of $\tilde b$ in theory B.

But we mentioned above that for fixed $M$ of rank $N_c$ and fixed sign of $B$ there could be either one or two vacua in theory A.  Are they mapped correctly?  For that we should be more careful about our gauge group.

When theory A is $Spin(N_c)$, for each VEV of $M$ of rank $N_c$ and each sign of $B$ there are also two choices for the monopole operator: $Y_{Spin}=0$ or $Y_{Spin}=\infty$ (related by $\Z_2^{\cal M}$).  In this case theory B is $SO(\tilde N_c)_-$, and we have
\eqn\vevtildey{{\tilde Y}^2={\tilde Y}_{Spin} = {\det_{N_c\times N_c}(M) \over \det(q q)} {\tilde Y}_{Spin,low} ={\det_{N_c\times N_c}(M) \over \det(q q)} \tilde Y^{(1)} \tilde Y^{(2)}= -{1 \over  {\tilde \eta}}~.}
The two possible values of $\tilde Y = \pm {i\over \sqrt{\tilde \eta}}$ are fixed points of the identification $\tilde Y \sim -{1\over \tilde \eta \tilde Y}$ \songauge, and hence lead to two different vacua.  They are related by the global $\Z_2^{\tilde {\cal M}}$ symmetry whose generator takes ${\tilde Y} \to -{\tilde Y}$, and are the dual of the
choice of $Y_{Spin}=0,\infty$ in theory A.  This interpretation is consistent with our mappings \firstmap.

When theory A is $SO(N_c)_-$, for each value of $M$ there are still two values of $B$, but there is a single choice, $Y=0$ (which is a fixed point of $\Z_2^{\cal M}$).  Correspondingly, theory B is $Spin (\tilde N_c)$ in which there are two values of $\tilde b$, as in \tildebBrel, but no additional freedom in the VEV of the monopole operator $\tilde Y_{Spin}$ in \vevtildey.

When theory A is $SO(N_c)_+$ there is freedom only in the sign of $B$ but not in $Y=0$.  In this case theory B is $SO({\tilde N}_c)_+$, and there are again two possible values for ${\tilde Y}$, which using \vevtildey\ are ${\tilde Y} = \pm i / \sqrt{\tilde \eta}$. But these two vacua are actually identified by \songauge, so we have a single vacuum, agreeing with the situation in theory A (and with \secondmap).

Thus, the mapping \sodu\ leads to a consistent mapping of all these vacua far on the Higgs branch in the $4d$ theory on a circle, using \firstmap\ and \secondmap.

\newsec{$3d$ dualities}

\subsec{Reduction of the $SO(N)_+$ duality to $3d$}
\subseclab\sondual

The duality we found up to now is not purely a $3d$ duality, since it involves the compact moduli spaces that we get in the $4d$ theory on a circle. In this subsection we will see how we can turn it into a bona fide duality of $3d$ gauge theories.

Consider the duality between $SO(N_c)_+$ and $SO({\tilde N}_c)_+$. The $SO(N_c)_+$ theory discussed above differs from the $3d$ $SO(N_c)$ theory by the superpotential \circlesuppot\ and by the compactness of the moduli space. Let us focus on theory A near the origin of the moduli space $Y=0$, and keep $|\eta Y^2| \ll 1.$
In the dual theory B, this means that we are (using \secondmap) near ${\tilde Y}=i/\sqrt{\tilde \eta}$ (or equivalently ${\tilde Y}=-i/\sqrt{\tilde \eta}$), with
$|\sqrt{\tilde \eta}\tilde Y - i| \ll 1$.

If we look at the low-energy superpotential in theory A we still have $W_A = \eta Z$, though the effects of this superpotential are very small in the region we are now discussing. In theory B we break the $SO({\tilde N}_c)$ theory at this value of ${\tilde Y}$ to $SO(N_f-N_c+2)\times SO(2)$.\foot{Naively, one may think that the symmetry is broken to
$S(O(N_f-N_c+2)\times O(2))$, with an extra $\Z_2$ factor. However, the
identification on the moduli space \songauge\ uses the Weyl
transformation $\sigma_1 \to -\sigma_1$, which is in $O(2)$, but
not in $SO(2)$. This means that the unbroken gauge symmetry in the
$SO({\tilde N}_c)_+$ theory around ${\tilde Y}=i/\sqrt{\tilde \eta}$ is
only $SO(N_f-N_c+2)\times SO(2)$.}
The operator ${\tilde Y}$ maps at low energies to the Coulomb branch coordinate $V_+$ of the $SO(2)$, and we can consider a new ${\tilde Y}_{low}$ Coulomb branch coordinate for the low-energy $SO(N_f-N_c+2)$ (defined as in \ysodef). In the low-energy superpotential of theory B we have contributions from the original $W_B$ of \circlesuppot.
The semi-classical forms of the monopole operators \zdef, \ysodef, imply that ${\tilde Z} = {\tilde Y} {\tilde Y}_{low}$. In addition we have an Affleck-Harvey-Witten superpotential related to the breaking of the $SO({\tilde N}_c)$ gauge group, which is proportional to ${\tilde Y}_{low} / {\tilde Y}$. Thus, the full low-energy superpotential near this point is
\eqn\wbeff{W_B = {1\over 2} M q q + {{\tilde Y}_{low} \over {\tilde Y}} + {\tilde \eta} {\tilde Y}_{low} {\tilde Y}.}

We can now use the mapping \secondmap\ between ${\tilde Y}$ and $Y$ to rewrite this in terms of $Y$, which is now an elementary field in theory B:
\eqn\wbefftwo{W_B = {1\over 2} M q q + 4 \sqrt{\eta {\tilde \eta}} {\tilde Y}_{low} {Y\over {1 + \eta Y^2}}.}
The choice of sign for the square root is arbitrary (the two choices are related by the global symmetry $Y\to -Y$).
We can now simply take $\eta \to 0$ on both sides (keeping $\eta {\tilde \eta} = (-1)^{N_f-N_c} / 256$ fixed); in theory A this is allowed since the effect
of the superpotential smoothly goes to zero in the region we are keeping, and in theory B the same is also true (since $|\eta Y^2 |\ll 1$). In this limit we find in theory A an $SO(N_c)$ $3d$ theory with a
non-compact Coulomb branch and with $W_A=0$, and in theory  B an $SO(N_f-N_c+2)$ $3d$ theory, again with a non-compact Coulomb branch, and with
\eqn\wbeffthree{W_B = {1\over 2} M q q + {i^{N_f-N_c}\over 4} {\tilde Y}_{low} Y,}
where $Y$ is now an elementary singlet in this low-energy theory, and ${\tilde Y}_{low}$ is its standard Coulomb branch coordinate \ysodef.

We can now lift this to a high-energy $3d$ duality between these two gauge theories,
by replacing ${\tilde Y}_{low}$ by the appropriate microscopic monopole operator ${\tilde y}$ of $SO(N_f-N_c+2)$, and the superpotential of theory B with
\eqn\wbeffthreen{W_B = {1\over 2} M q q + {i^{N_f-N_c}\over 4} {\tilde y}\, Y~.}
Note that unlike in other cases discussed in \ARSW, here we did not need to perform any real mass deformation in order to obtain the duality for the standard $3d$ SQCD theory from $4d$, but just to take the $3d$ limit carefully. In the $3d$ limit we have an extra global $U(1)_A$ symmetry, that was anomalous in $4d$. The quantum numbers of the various operators are consistent with the superpotential \wbeffthreen; using a specific choice for the $3d$ R-symmetry, they are
\eqn\Ahaethree{
\vbox{\offinterlineskip\tabskip=0pt
\halign{\strut\vrule#
&~$#$~\hfil\vrule
&~$#$~\hfil\vrule
&~$#$~\hfil
&~$#$~\hfil
&~$#$\hfil
&~$#$\hfil
&~$#$\hfil
&\vrule#
\cr
\noalign{\hrule}
&  &  SO(N_c) & SU(N_f) & U(1)_A & U(1)_R &\quad \Z_2^\CC\quad &\quad \Z_2^{\cal M}\quad &\cr
\noalign{\hrule}
&  Q         & \; {\bf  N_c}     & \; {\bf N_f}    &    \quad  1 &  \quad  0   & & &\cr
\noalign{\hrule}
& M              & \; {\bf  1}     & \; {\bf N_f(N_f+1)/2}   &    \quad  2 &  \quad  0   & \quad +1 &\quad  +1&\cr
& Y              & \; {\bf  1}     & \; {\bf 1}   &    \quad -N_f &  \quad  N_f-N_c+2 \qquad  &\quad +1& \quad -1&\cr
&B              & \; {\bf  1}     & \; ({\bf N_f})^{N_c}_A   &    \quad N_c &  \quad  0 \qquad &\quad -1&\quad  +1&\cr
& \beta              & \; {\bf  1}     & \;({\bf N_f})^{N_c-2}_A   &    \quad N_c-N_f-2 &   \quad  N_f-N_c+2    &\quad -1&\quad  -1&\cr }
\hrule}}
in theory A, and
\eqn\Ahamthree{
\vbox{\offinterlineskip\tabskip=0pt
\halign{\strut\vrule#
&~$#$~\hfil\vrule
&~$#$~\hfil\vrule
&~$#$~\hfil
&~$#$~\hfil
&~$#$\hfil
&~$#$\hfil
&~$#$\hfil
&\vrule#
\cr
\noalign{\hrule}
&  &  SO(N_f-N_c+2) & SU(N_f) & U(1)_A & U(1)_R &\quad \Z_2^\CC\quad &\quad \Z_2^{\tilde {\cal M}}\quad &\cr
\noalign{\hrule}
&  q         & \; {\bf  N_f-N_c+2}     & \; {\bf {\overline N_f}}    &    \quad  -1 &  \quad  1   & & &\cr
& M              & \; {\bf  1}     & \; {\bf N_f(N_f+1)/2}   &    \quad  2 &  \quad  0   &\quad  +1 &\quad +1&\cr
& Y              & \; {\bf  1}     & \; {\bf 1}   &    \quad -N_f &  \quad  N_f-N_c+2 &\quad +1&\quad -1&\cr
\noalign{\hrule}
& {\tilde y}       & \; {\bf  1}     & \; {\bf 1}   &    \quad N_f &  \quad  N_c-N_f \qquad  & \quad +1&\quad -1&\cr
&\tilde B            & \; {\bf  1}     & \; ({\bf N_f})^{N_c-2}_A   &    \quad N_c-N_f-2 &  \quad  N_f-N_c+2    &\quad -1&\quad +1&\cr
&\tilde \beta             & \; {\bf  1}     & \; ({\bf N_f})^{N_c}_A  &    \quad N_c &  \quad  0  &\quad -1&\quad -1&\cr
}
\hrule}}
in theory B. Note that $SU(N_f)\times U(1)_A$ is really $U(N_f)$. $({\bf N_f})^{N_c-2}_A$ and $({\bf N_f})^{N_c}_A$ denote totally antisymmetric products. $\Z_2^\CC$ is the charge conjugation symmetry, generated by $\CC$, and $\Z_2^{\cal M}$ and $\Z_2^{\tilde {\cal M}}$ are the global symmetries \globalsymm, generated by ${\cal M}$ and ${\tilde {\cal M}}$, respectively.  We included their action only on the gauge singlets. The composites ${\tilde B}$ and ${\tilde \beta}$ in theory B are defined as in theory A (see \betadef), and their identification in theory A will be discussed below.

The three symmetries $U(1)_A$, $\Z_2^\CC$ and $\Z_2^{\cal M}$ are actually not independent. In theory A with gauge group $SO(N_c)$, the action of $e^{\pi i A}$ (which is in $SU(N_f)$ for even values of $N_f$) on $Q$ is part of the gauge group for even values of $N_c$, and is the same as $\CC$ for odd values of $N_c$. The action of $e^{\pi i A}$ on $Y$ is the same as ${\cal M}^{N_f}$. Thus, on gauge-invariant operators we have
\eqn\symmrel{e^{\pi i A} \CC^{N_c} {\cal M}^{N_f} = 1,}
and $SU(N_f)\times U(1)_A\times \Z_2^{\CC}\times \Z_2^{\cal M}$ is really $(U(N_f)\times \Z_2^{\CC}\times \Z_2^{\cal M})/\Z_2$. In the dual theory we have $e^{\pi i A} \CC^{N_f-N_c+2} {\tilde {\cal M}}^{N_f} = 1$, implying that for odd values of $N_f$,
\eqn\symmmap{{\cal M} \longleftrightarrow {\tilde {\cal M}} \CC.}
We will see below that this must be true for even values of $N_f$ as well.

The duality we find is very similar to the one discussed for $O(N_c)$ theories (more precisely, $O(N_c)_+$ theories) in \refs{\BeniniMF,\HwangHT,\AharonyCI}. Indeed, if we now gauge the charge conjugation symmetry $\Z_2^{\cal C}$ on both sides, we obtain precisely that duality, so our discussion is a derivation of this duality from $4d$. But we obtain a duality also for $SO(N_c)$ groups, meaning that there should be a consistent mapping of the charge-conjugation-odd baryons between the two sides. We can follow what happens to the $4d$ baryon mapping by our reduction procedure. In $4d$ the baryon $B=Q^{N_c}$ mapped to ${\tilde b}_{4d}/2{\sqrt {\tilde \eta}}$ \tildebBrel. In the reduction on the circle we say that this first becomes equal to $\sqrt{{\tilde \eta}} {\tilde b} / 2$ \tildebBrel, where the latter operator \bdef\ involves a monopole operator in $so(4)$. When we go onto the Coulomb branch as above, this monopole operator ${\tilde Y}_-$ becomes $i/\sqrt{\tilde \eta}$ (from the VEV of ${\tilde Y}$) times the odd monopole operator ${\tilde W}_-$ of $SO(2)$, so we find that $B$ maps to $i {\tilde \beta} / 2$, with ${\tilde \beta}$ defined as in \betadef. This is consistent with their global symmetry quantum numbers as in \Ahaethree, \Ahamthree.

The $4d$ operator $b_{4d}$ goes to zero in the $\eta \to 0$ limit that we described, as does its $4d$ dual ${\tilde B}_{4d}=q^{N_f-N_c+4}$ (since only $(N_f-N_c+2)$ components of the quarks remain massless in the limit we took in theory B). However, we now get a new relation (required by consistency of the duality), mapping $\beta$ to the $3d$ baryon ${\tilde B}=q^{N_f-N_c+2}$. We cannot derive this duality directly from $4d$, but on the part of the moduli space where we break both gauge groups to $SO(2)$, it follows by dualizing the vector multiplets \baryonsfourd\ in the $4d$ relation ${\cal W}_{\alpha} \leftrightarrow {\tilde {\cal W}}_{\alpha}$ into chiral multiplets (taking into account again the VEV of ${\tilde Y}$ in theory B). It is also consistent with the global symmetries as in \Ahaethree, \Ahamthree.  We conclude that the baryons map in the $3d$ duality between $SO(N_c)$ and $SO(N_f-N_c+2)$ by
\eqn\baryonmap{\{B, \beta\} \qquad \longleftrightarrow \qquad \{{i\over 2}{\tilde \beta}, -2i{\tilde B}\}.}
Note that this mapping requires that the $\Z_2^{\cal M}$ symmetry \globalsymm, which takes $B\to B$ and $\beta \to -\beta$, maps under the duality by \symmmap\ for all values of $N_f$.
As we mentioned above, in the $3d$ theory $b$ and $Z$ are not chiral, so they do not have a simple mapping under the duality.

We can perform many tests of this duality, comparing moduli spaces, chiral operators, deformations, and so on, but most of these tests are identical to tests of the $O(N_c)_+$ duality that were already performed in \AharonyCI. We can find new tests by involving also the baryon operators. For instance, suppose that
we turn on a VEV for $M$ of rank $N_c$, as in our discussion of the previous section. In theory A we still have two vacua for every such $M$, with $B^2 = \det_{N_c\times N_c}(M)$. In theory B we now give a mass to $N_c$ quarks, so that we are left at low energies with $N_f-N_c={\tilde N}_c-2$ massless quarks, and with a low-energy Coulomb branch coordinate ${\tilde Y}_{N_f-N_c} = {\tilde y} / \sqrt{\det_{N_c\times N_c}(M)}$. Our discussion around \betarel\ implies that in this low-energy theory, in which the superpotential sets its meson $qq=0$, there is a relation ${\tilde \beta}_{N_f-N_c}^2 = -4$. Translating this into the high-energy theory (using \massrel) we find ${\tilde \beta}^2 = - 4 \, \det_{N_c\times N_c}(M)$, so that we can indeed identify ${\tilde \beta}$ with $(- 2 i B)$.

For low values of $N_c$ and of $N_f-N_c$ there are slight modifications of this discussion, as in the $4d$ duality \IntriligatorID\ and in the analysis of the pure $3d$ theory \AharonyCI, but these do not raise any new issues so we will not discuss them in detail here.

\subsec{Reduction of the $Spin(N_c) \leftrightarrow SO({\tilde N}_c)_-$ duality to $3d$}

We can similarly obtain the dual of the $3d$ $Spin(N_c)$ theory, by starting from the $4d$ duality between $Spin(N_c)$ and $SO({\tilde N}_c)_-$. We can again focus on the same points $Y_{Spin}=0$ and ${\tilde Y}=-i/\sqrt{\tilde \eta}$ in the moduli space, and obtain the low-energy superpotential \wbeff. However, now we are at a fixed point of \songauge, so the discussion in footnote 9 implies that the unbroken gauge symmetry in theory B is $S(O(N_f-N_c+2)\times O(2))$.  The moduli space coordinate $\tilde Y$ is the monopole operator of the $SO(2)$ factor; when expressed in terms of $\tilde Y$, the $SO(2) \subset O(2)$ is not visible.  Focusing on the region around ${\tilde Y}=-i/\sqrt{\tilde \eta}$, it is convenient to change variables from the approximate free field $\tilde Y$ to $y$ defined by
\eqn\ydef{4 \sqrt{\eta {\tilde \eta}} {y\over {1 + \eta y^2}} = {1\over {\tilde Y}} + {\tilde \eta} {\tilde Y}~.}
Now we expand around $y=0$, the mapping from theory A to theory B \firstmap\ identifies $Y_{Spin} = y^2$, and the extra $\Z_2$ gauge identification acts as $y \to -y$. This $\Z_2$ also acts on the $SO(N_f-N_c+2)$ theory as charge conjugation, and it also changes the sign of the monopole operator ${\tilde y} ={\tilde Y}_{low}$ of $SO(N_f-N_c+2)$. Thus, we recognize the gauge theory we get as an $O(N_f-N_c+2)_-$ theory, with the elementary field $y$ odd under $\Z_2^{\tilde \CC} \subset O(N_f-N_c+2)$.  The low energy superpotential after taking $\eta \to 0$ is
\eqn\wbeffthreens{W_B = {1\over 2} M q q + {i^{N_f-N_c}\over 4} {\tilde y}\, y~.}
The bottom line is that the dual of the $Spin(N_c)$ SQCD theory is similar to \wbeffthreen, but the dual gauge group is $O(N_f-N_c+2)_-$, and $y$ and $\tilde y$ are odd under its $\Z_2^{\tilde \CC}$ subgroup. 

The $\Z_2^\CC$ global symmetry of theory A is mapped under this duality to $\Z_2^{\tilde \MM}$. In the $Spin(N_c)$ theory $B$ is a $\Z_2^\CC$-odd gauge-invariant operator, while $Y$ and $\beta$ do not exist. In the dual $O(N_f-N_c+2)_-$ theory ${\tilde \beta}$ is a $\Z_2^{\tilde \MM}$-odd gauge-invariant operator (mapped to $B$), while $y$, ${\tilde y}$ and ${\tilde B}$ are not gauge-invariant. The operator $\beta_{Spin}$ of \betaspindef\ is present in theory A, and maps to the gauge-invariant operator ${\tilde B}y$ in theory B.

In the discussion above we expanded around the point $Y_{Spin}=0$, which is mapped to $\tilde Y = -i/\sqrt{\tilde \eta}$.  Instead we could expand around $\tilde Y=0$, which corresponds to $Y_{Spin}= -1/\eta$.  Here the $SO(N_f-N_c+4)$ gauge group of theory B is unbroken, while the gauge group of theory A is broken as $Spin(N_c) \to (Spin(N_c-2) \times Spin(2))/\Z_2$, where the $\Z_2$ acts on the spinors in both groups.  Next we analyze the low energy dynamics of theory A, as we did in theory B above, focusing on the $Spin(2)$ dynamics.  It is important that there are no massless fields charged under this group.  Hence, its Wilson loops become trivial at long distances.  Further, we can map each $(Spin(N_c-2) \times Spin(2))/\Z_2$ bundle to an $SO(N_c-2)$ bundle by simply ignoring the $Spin(2)$ transition functions.  Hence, when we integrate out the $Spin(2)$ dynamics we are left with an $SO(N_c-2)$ gauge theory.  The dual of the $Spin(2)$ gauge field, $Y_{Spin}$, is an almost free field, which we can replace using \firstmap\ by an elementary field $\tilde Y$, identified with the monopole operator of theory B.  Finally, we can identify the gauge-invariant operator $(i Y_{low,Spin}/ \sqrt{\eta} Z)$ in theory A (where $Y_{low,Spin}$ is the Coulomb branch coordinate of $Spin(N_c-2)$) with $Y_{low}$ of $SO(N_c-2)$, and, 
as in  our analysis above, the various monopole-instantons couple $\tilde Y$ to $Y_{low}$.  For $\tilde \eta \to 0$ we find a superpotential term proportional to $Y_{low} \tilde Y$, as in \wbeffthree.  The duality we derive this way is precisely the inverse of the $SO(N)$ duality that we derived in section \sondual. This is a non-trivial consistency check on our web of dualities, because in section \sondual\ we derived this duality from the compactification of a different $4d$ duality.

We can also obtain a dual for $Pin(N_c)$, by gauging the global symmetry $\Z_2^{\cal M}$ in the duality for $O(N_c)_+$ groups. The fact that in the $O(N_c)_+$ duality the symmetry $\Z_2^{\cal M}$ maps to itself implies that the $Pin(N_c)$ theory is dual to a $Pin(N_f-N_c+2)$ theory.

\subsec{Dualities with Chern-Simons terms}
\subseclab\csdual

As in \ARSW, we can flow from the duality above to a duality with Chern-Simons terms. We can obtain an $SO(N_c)$ theory with $N_f$ flavors and a Chern-Simons term at level $k > 0$ by starting from the theory with $N_f+k$ flavors and giving $k$ flavors a positive real mass, by turning on a background field for the $U(N_f+k)$ global symmetry. In the dual $SO(N_f+k-N_c+2)$ theory, this maps to giving $k$ flavors a negative real mass, giving the mesons they couple to a positive real mass, and also giving a negative real mass to the singlet $Y$. Integrating out the massive fields we find an $SO(N_f+k-N_c+2)$ theory with $N_f$ flavors, level $(-k)$, and a $W_B = {1\over 2} M q q$ superpotential. This is precisely the duality conjectured in \KapustinGH\ for the $O(N)$ theories (more precisely, $O(N)_+$ theories), and here we see that it is true also for $SO(N)$.

The difference between $O(N)_+$ and $SO(N)$ is that now we need to understand also how to map the baryon operators in the two sides, and this is more complicated (as in the discussion of $SU(N_c)$ Chern-Simons-matter theories in \ARSW) since they involve monopoles. In theory A we still have the baryon operator $B = Q^{N_c}$, while the baryon-monopole $\beta = Q^{N_c-2} W_-$ is no longer gauge-invariant in the presence of the Chern-Simons term, and similarly in theory B. We claim, similar to what we found for $SU(N_c)$ Chern-Simons-matter theories, that the dual of $B$ is now given by a monopole operator ${\tilde \beta}' = q^{N_f-N_c} {\tilde W}_- ({\tilde W}_{\alpha})^k$, which is gauge-invariant. The quarks are contracted with an epsilon symbol, and break $SO(N_f+k-N_c+2)$ to $SO(k+2)$. The monopole operator ${\tilde W}_-$ breaks $SO(k+2)$ to $SO(k)\times U(1)$, and because of the Chern-Simons term it carries a charge $(-k)$ under the $U(1)$. The gluinos ${\tilde W}_{\alpha}$ are off-diagonal gluinos in $SO(k)\times U(1)$ which cancel this charge, and carry $k$ different vector indices of $SO(k)$ that are contracted by an epsilon symbol, such that the total operator is gauge-invariant. One can verify that the global symmetry charges of this $\tilde \beta'$ exactly match with those of $B$. Similarly, we can construct an operator $\beta'$ in theory A that matches with ${\tilde B} = q^{N_f+k-N_c+2}$.

In theory A we have the relation $B^2 = \det_{N_c\times N_c} (M)$. To see this in theory B we turn on a VEV of rank $N_c$ for $M$, leaving $N_f-N_c$ massless flavors $q$, and we then ignore for a moment the superpotential and imagine giving an expectation value to the remaining massless flavors. This breaks the gauge group to $SO(k+2)$ with level $(-k)$ and no massless flavors. At low energies this is a purely topological theory, in which we can construct a singlet operator ${\tilde \beta}' = {\tilde W}_- {\tilde W_{\alpha}}^k$ as above, which is charged under the charge conjugation symmetry of this theory, and argue that it squares to one (similar to our discussion of the $SU(k)$ theory at level $(-k)$ in \ARSW). Lifting this to the high-energy theory using \massrel\ we get precisely the expected relation (which turns out to be independent of the VEVs of the $q$'s, so it is valid even for $q=0$).

For $N_f=0$ our duality reduces to a duality of pure supersymmetric Chern-Simons theories, $SO(N_c)_k$ being identified with $SO(|k|-N_c+2)_{-k}$. This is just the standard level-rank duality of $SO(N)$ Chern-Simons theories. At low energies we can integrate out the gauginos, shifting the $SO(N_c)$ level to $k \to k - (N_c-2) {\rm sign}(k)$. We then obtain the standard level-rank relation \refs{\Naculich,\Mlawer,\NakanishiHJ}
\eqn\levelrank{SO(n)_m \longleftrightarrow SO(m)_{-n},}
for $n,m > 0$, that can be proven by studying $nm$ real fermions in two dimensions.

Similarly, we can flow from our $Spin(N_c)$ duality to find a duality between $Spin(N_c)_k$ and $(O(N_f+|k|-N_c+2)_-)_{-k}$ (and a corresponding non-supersymmetric level-rank duality taking $Spin(n)_m$ to $(O(m)_-)_{-n}$), and a duality between the supersymmetric $Pin(N_c)_k$ and $Pin(N_f+|k|-N_c+2)_{-k}$ theories (with a level-rank duality taking $Pin(n)_m$ to $Pin(m)_{-n}$).

\

\subsec{The special case of $SO(2)=U(1)$ with $N_f=2$: a triality of dualities}

We now have two different dualities for $SO(2)=U(1)$ gauge theories with $N_f$ flavors, which we refer to as theory A. First, we can view the gauge group as $U(1)$ and find a dual theory based on $U(N_f-1)$ \AharonyGP.  We will refer to this dual theory as B1.  Alternatively, as in this paper, we can view it as $SO(2)$ and find a dual theory based on $SO(N_f)$.  We will refer to it as B2.  The B1 dual exhibits the full $SU(N_f)\times SU(N_f)\times U(1)_A\times U(1)_J \times U(1)_R$ global symmetry, while the B2 dual exhibits explicitly only $SU(N_f)\times U(1)_A \times U(1)_R$, and the other symmetries arise as accidental symmetries at low energies.

The $SO(2)$ theory with $N_f=2$ deserves special attention.  In this case the gauge groups of theories A, B1 and B2 are all $U(1)$, and they all have two flavors.  Furthermore, in this case there is also a mirror theory, that also has gauge group $U(1)$ and two flavors \refs{\IntriligatorEX\deBoerMP-\deBoerKA,\AharonyBX}.  We will refer to this theory as B3.

Let us understand the relation between these dual descriptions (see also \AharonyCI).
We begin with theory A.  We can think of it either as a $U(1)$ theory with two flavors $Q^a$, $\tilde Q^{\tilde a}$ ($a,{\tilde a}=1,2$), or as an $SO(2)$ theory with two doublets $P_i$ ($i=1,2$).
Let us work out the translation between these two languages. In the $U(1)$ description of this theory,
the chiral operators are the magnetic monopoles
$V_\pm$ \sotwocoords\ and four mesons ${\hat M}^{a \tilde b}=Q^a{\tilde Q}^{\tilde b}$. The translation to $SO(2)$ variables, if we keep the standard normalization for the kinetic terms, is by
\eqn\uonesotwo{Q^i = {1\over \sqrt{2}} (P_i^1 + i P_i^2),\qquad
{\tilde Q}^i = {1\over \sqrt{2}} (P_i^1 - i P_i^2)\,.}
Defining the standard $SO(2)$ mesons $M_{ij} = P_i \cdot P_j$, the symmetric part of ${\hat M}$ is related to $M$ by ${\hat M}_{ij} + {\hat M}_{ji} = M_{ij}$. The anti-symmetric part of ${\hat M}$ is related to the $SO(2)$ baryon $B \equiv P_1^1 P_2^2 - P_1^2 P_2^1$ by ${\hat M}_{12} - {\hat M}_{21} = - i B$.
The natural monopole-related operators in the $SO(2)$ language are
\eqn\vstoys{
Y\equiv V_++V_-~,\qquad\qquad
\beta \equiv V_+-V_-\,.
}
The former is the basic monopole in the $SO(2)$ language, and the latter is the baryon operator $\beta$ \betadef\ in this special case.
The charges of the different objects under the global $U(1)_A\times U(1)_R$ symmetry that is visible in all descriptions are (using our standard conventions):
\eqn\chargesA{
\vbox{\offinterlineskip\tabskip=0pt
\halign{\strut\vrule#
&~$#$~\hfil\vrule
&~$#$~\hfil\vrule
&~$#$~\hfil
&\vrule#
\cr
\noalign{\hrule}
&  &   U(1)_A & U(1)_R &\cr
\noalign{\hrule}
&  Q ~, ~\tilde Q~,~P     & \; 1     & \; 0    &   \cr
& Y ~, ~\beta        & \; -2    & \; 2  &   \cr
&B ~, ~M           & \; 2     & \; 0   &  \cr
}
\hrule}}

The dual description which has all the symmetries of theory A manifest is B1. This is a $U(1)$ theory with two flavors $q^{(1)}_a$, ${\tilde q}^{(1)}_{\tilde a}$ ($a,{\tilde a}=1,2$) and with additional singlet fields ${\hat M}^{a{\tilde b}}_{(1)}$ and $V_\pm^{(1)}$. The superpotential is \AharonyGP\
\eqn\supSObb{
W_{B1}=q^{(1)}\,{\hat M}^{(1)}\,\tilde q^{(1)}+V_+^{(1)}\tilde V_-^{(1)}+V_-^{(1)}\tilde V_+^{(1)}\,.
}
For comparison with other duals it will be useful to translate this to the $SO(2)$ language, as
we did above. We define as above
\eqn\chiralsbb{
\tilde Y^{(1)}\equiv \tilde V_+^{(1)}+\tilde V_-^{(1)}\,,\qquad
\tilde \beta^{(1)}\equiv \tilde V_+^{(1)}-\tilde V_-^{(1)}\,,\qquad
\tilde B^{(1)} \equiv i (q_1^{(1)} {\tilde q}_2^{(1)} - q_2^{(1)} {\tilde q}_1^{(1)})\,.
}
Translating the quarks $q,{\tilde q}$ to $SO(2)$ quarks $p$ as above, the superpotential \supSObb\ becomes
\eqn\supSObbtwo{
W_{B1}= {1\over 4} M_{ab}^{(1)}\, p^{(1)}_a\,p^{(1)}_b- {1\over 2}
B^{(1)}\,\tilde B^{(1)}+{1\over 2}Y^{(1)} \,\tilde Y^{(1)}-{1\over 2}\beta^{(1)} \tilde \beta^{(1)}\,,
}
with the singlets $B^{(1)}$ and $M^{(1)}$ related to ${\hat M}^{(1)}$, and the singlets $\beta^{(1)}$ and $Y^{(1)}$ related to $V_\pm^{(1)}$, as in theory A. These singlets are identified with the corresponding operators in theory A.
The $U(1)_A\times U(1)_R$ charges of the different objects are:
\eqn\chargesBbb{\vbox{\offinterlineskip\tabskip=0pt
\halign{\strut\vrule#
&~$#$~\hfil\vrule
&~$#$~\hfil\vrule
&~$#$~\hfil
&\vrule#
\cr
\noalign{\hrule}
&  &   U(1)_A & U(1)_R &\cr
\noalign{\hrule}
&  q^{(1)} ~, ~\tilde q^{(1)}~,~p    & \; -1     & \; 1    &   \cr
& \tilde Y^{(1)} ~, ~\tilde \beta^{(1)}        & \; 2    & \; 0 &   \cr
&\tilde B^{(1)}            & \; -2     & \; 2   &  \cr
}
\hrule}}

The normalization of the first term in \supSObbtwo\ is half of the normalization in the standard $SO(N_c)$ duality \wbeffthreen, and it will be easier to compare them if we have the same normalization in both cases. Since the $p$'s do not appear in any gauge-invariant chiral operator, we can simply rescale them to new variables ${\hat p}^{(1)}_i = p^{(1)}_i / \sqrt{2}$. This also rescales the baryon ${\tilde B}^{(1)}$ to ${\tilde {\hat B}}^{(1)} = {\tilde B}^{(1)} / 2$, and because of the relation of the quantum numbers of the monopoles to those of the quarks, the latter are also rescaled to ${\tilde {\hat Y}}^{(1)} = 2 {\tilde Y}^{(1)}$, ${\tilde {\hat \beta}}^{(1)} = 2 {\tilde \beta}^{(1)}$. We can now write \supSObbtwo\ as
\eqn\supSObbtworescaled{
W_{B1}= {1\over 2} M_{ab}^{(1)}\, {\hat p}^{(1)}_a\,{\hat p}^{(1)}_b-
B^{(1)}\,{\tilde {\hat B}}^{(1)}+{1\over 4}Y^{(1)} \,{\tilde {\hat Y}}^{(1)}-{1\over 4}\beta^{(1)} {\tilde {\hat \beta}}^{(1)}\,.
}

The dual description B2, with gauge group $SO(2)$, is quite similar to B1. The difference is that we do not have the singlet fields $B$ and $\beta$, and the superpotential is \wbeffthreen
\eqn\supSObatwo{
W_{B2}= {1\over 2}M_{ab}^{(2)}\, p^{(2)}_a \,p^{(2)}_b+{1\over 4} Y^{(2)}\,\tilde Y^{(2)}\,.
}
The map between the chiral operators here is
\eqn\mapba{
M\to M^{(2)}\,,\qquad B\to {i\over 2} \tilde \beta^{(2)}\,,\qquad \beta\to -2i \tilde B^{(2)}\,,\qquad Y\to Y^{(2)}\,.
}
Note that in description B2 the $U(1)_J$ symmetry is not present in the UV gauge theory, as the singlet $Y$ mixes with $\beta$ under this symmetry.
Moreover only an $SU(2) \subset SU(2)\times SU(2)$ flavor symmetry is visible in the UV. The symmetries broken in this UV description should appear as accidental symmetries of the IR physics.

Finally, in description B3 we do not have $M_{12}$ and $Y$, and the superpotential is
\eqn\supSObctwo{
W_{B3}= {\hat M}^{11}_{(3)}\, q^{(3)}_{1 }\,{\tilde q}^{(3)}_{1}+ {\hat M}^{22}_{(3)}\, q^{(3)}_{2 }\,{\tilde q}^{(3)}_{2}\, = {1\over 4} M_{11}^{(3)} p_1^{(3)} p_1^{(3)} + {1\over 4} M_{22}^{(3)} p_2^{(3)} p_2^{(3)}.
}
Using the fact that mirror symmetry exchanges the monopoles $V_{\pm}$ with the off-diagonal mesons ${\hat M}^{12}, {\hat M}^{21}$ with coefficient one, the map between the chiral operators here is
\eqn\mapbc{
M_{\ell\ell}\to M_{\ell\ell}^{(3)}\,,\qquad B\to i \tilde \beta^{(3)}\,,\qquad \beta\to -i \tilde B^{(3)}\,,\qquad Y\to p^{(3)}_{1}\,p^{(3)}_{2}\,,\qquad M_{12} \to \tilde Y^{(3)}\,.
}
Here $U(1)_J$ is present in the UV description and is identified with part of the dual flavor group, but only $U(1) \times U(1) \subset SU(2)\times SU(2)$ is a symmetry of the UV theory. As in the B1 theory, it is convenient to rescale the $p$'s by $\sqrt{2}$, and in the rescaled variables defined as above we have
\eqn\supSObctworescaled{
W_{B3}= {1\over 2} M_{11}^{(3)} {\hat p}_1^{(3)} {\hat p}_1^{(3)} + {1\over 2} M_{22}^{(3)} {\hat p}_2^{(3)} {\hat p}_2^{(3)},
}
with the new mapping
\eqn\mapbcrescaled{
M_{\ell\ell}\to M_{\ell\ell}^{(3)}\,,\qquad B\to {i\over 2} {\tilde {\hat \beta}}^{(3)}\,,\qquad \beta\to -2 i {\tilde {\hat B}}^{(3)}\,,\qquad Y\to 2 {\hat p}^{(3)}_{1}\,{\hat p}^{(3)}_{2}\,,\qquad M_{12} \to {1\over 2} {\tilde {\hat Y}}^{(3)}\,.
}
Note that the mapping between the $B$'s and $\beta$'s is now the same as in \mapba.

Let us now relate these theories, by understanding their deformations.  We claim that the IR superconformal field theory that all these theories flow to has eight marginal deformations, and that two of them are exactly marginal.  In the B1 description the eight marginal operators are  ${\hat M}^{a{\tilde a}}_{(1)} V_\pm^{(1)}$.  Denote their coupling constants by $\lambda_{a {\tilde a}}^{\pm} $.
The space of exactly marginal deformations is generally given by the (complexified)
quotient of the space of marginal deformations by the global symmetries, which act on it non-trivially \GreenDA.  This can be found by noting the invariant combinations constructed out of $\lambda_{a {\tilde a}}^{\pm} $.  All these deformations preserve $U(1)_A$, but they are charged under $SU(2)\times SU(2)\times U(1)_J$.  There are two non-trivial invariants of this group: $\epsilon^{ab} \epsilon^{{\tilde a}{\tilde b}} \lambda_{a {\tilde a}}^+ \lambda_{b {\tilde b}}^-$ and $\left(\epsilon^{ab} \epsilon^{{\tilde a}{\tilde b}} \lambda_{a {\tilde a}}^+ \lambda_{b {\tilde b}}^+\right)\left(\epsilon^{cd} \epsilon^{{\tilde c}{\tilde d}} \lambda_{c {\tilde c}}^- \lambda_{d {\tilde d}}^-\right)$.
This shows that there are two exactly marginal deformations.  Equivalently, the global symmetry $SU(2)\times SU(2)\times U(1)_J$ is not completely broken on the space of couplings, but a $U(1)$ always remains (which is a subgroup of the diagonal $SU(2)$).

By a global symmetry transformation, we can choose the two exactly marginal deformations to be given by (in the $SO(2)$ language)
\eqn\exactlymarginal{
\delta W=\gamma\,B\,\beta+\rho\, M_{12}\,Y\,.
}
Note that these deformations are invariant under $\Z_2^\CC$, but they break $\Z_2^{\cal M}$ \globalsymm.

Suppose we first add the term with $\gamma$ to theory A. In the B1 dual description, the superpotential becomes after integrating out the massive singlet fields
\eqn\supSObbtwo{
W_{B1}+\gamma\,B^{(1)}\,\beta^{(1)} \rightarrow
{1\over 2} M_{ab}^{(1)}\, {\hat p}^{(1)}_a\,{\hat p}^{(1)}_b +
{1\over 4} Y\,{\tilde {\hat Y}}^{(1)}-{1\over {4\gamma}}\,{\tilde {\hat \beta}}^{(1)}\,{\tilde {\hat B}}^{(1)}.
}
This is exactly the same as the superpotential $W_{B2}$, deformed by ${-1\over {4\gamma}}\,\tilde B^{(2)}\,\tilde \beta^{(2)}$.
On the other hand, adding $\gamma\,B\,\beta$ to theory A translates in theory B2 to adding $\gamma\,\tilde B^{(2)}\,\tilde \beta^{(2)}$. These two descriptions should be equivalent, so we conclude that our deformed IR superconformal field theory possesses an exact duality under $\gamma\longleftrightarrow (-1/4\gamma)$.

Next we add the term with $\rho$ to theory A. The superpotential in the dual B2 description becomes, after integrating out massive singlet fields,
\eqn\supSObbtwo{
W_{B2}+\rho\, M_{12}^{(2)}\,Y^{(2)} \rightarrow {1\over 2} M_{11}^{(2)}\, p^{(2)}_{1 }\,p^{(2)}_{1}+ {1\over 2} M_{22}^{(2)}\, p^{(2)}_{2 }\,p^{(2)}_{2}-{1\over {4\rho}}\,\tilde Y^{(2)}\,p^{(2)}_{1}\,p^{(2)}_{2}\,.
}
This is exactly the same as the superpotential $W_{B3}$, deformed by $-{1\over {4\rho}}\,{\tilde {\hat Y}}^{(3)}\,{\hat p}^{(3)}_{1}\,{\hat p}^{(3)}_{2}$; as we mentioned, the baryon mappings are also consistent. On the other hand, adding $\rho\, M_{12}\,Y$ to theory A translates in theory B3 to adding $\rho\,{\tilde {\hat Y}}^{(3)}\,{\hat p}^{(3)}_{1}\,{\hat p}^{(3)}_{2}$.
Thus we conclude that our IR superconformal field theory also possesses an exact duality $\rho\longleftrightarrow (-1/4\rho)$.

Assuming the three dualities, we deduced that the exactly marginal
couplings enjoy dualities taking $\gamma\longleftrightarrow (-1/4\gamma)$ and
$\rho\longleftrightarrow (-1/4\rho)$. Alternatively, if one could prove the duality of the marginal deformations,
one could deduce all three duals of $U(1)$ with two flavors from knowing any one of them.

\

\

\newsec{Partition functions and indices for $so(N)$ dualities}

A set of useful  checks of  dualities is given by  comparisons
of supersymmetric partition functions of the putative dual pair:
if the two $3d$ UV theories describe the same IR physics,
their $\S^3$ and $\S^2\times \S^1$ supersymmetric partition
functions should agree.
In this section we will discuss these checks for the dualities of the previous section.

\

\subsec{The  partition function on $\S^2\times \S^1$}

Let us start by discussing the matching of the partition function on $\S^2 \times \S^1$, also known as the supersymmetric index.
The indices for the $O(N_c)_+$
versions of the dualities discussed in this paper were checked to match in \refs{\HwangQT,\HwangHT}.
The index is sensitive to the global structure of the gauge group,
and thus the matching of the indices for $SO(N_c)$ does not directly follow from these computations.\foot{
We will discuss below a generalization of the $O(N_c)_+$ index, which contains the same information as the $SO(N_c)$ index.}
We will check here that the supersymmetric indices match also for the $SO(N_c)$ dualities.  In the process of doing this, we will see that, since the index contains information about local operators, it can test the proposed mapping of the baryon operators to the baryon-monopole operators discussed in the
preceding sections.

First, let us briefly review the definition of the $3d$ supersymmetric index. It is defined by the following trace over states on $\S^2 \times \R$ (see~\refs{\BhattacharyaZY\KimWB\ImamuraSU\KapustinJM-\DimoftePY,\ARSW} for details):
\eqn\indexDefthree{
\II(x;\,\{u_a\})=\Tr \left[(-1)^{2J_3}\,x^{\Delta+J_3}\,\prod_{a}u_a^{e_a}\right]\,.
}
Here $\Delta$ is the energy in units of the $\S^2$ radius (related to the conformal dimension for superconformal field theories), $J_3$ is the Cartan generator of the Lorentz $SO(3)$ isometry of $\S^2$, and $e_a$ are charges under $U(1)$ global symmetries (which could be subgroups of non-Abelian global symmetries). The states that contribute to this index satisfy $\Delta-R-J_3=0$,
where $R$ is the R-charge (that is used in the compactification on $\S^2$).

This index can be computed by a partition function on $\S^2 \times \S^1$, and localization dictates that the index gets contributions only from BPS configurations. For example, for a $U(1)$ gauge multiplet, we can take the gauge field to have a holonomy $z \in U(1)$ around the $\S^1$ and magnetic flux $m \in \Z$ on the $\S^2$, which then determines the configurations of the other fields in the gauge multiplet.  The $1$-loop determinant of a chiral multiplet of R-charge $R$ coupled with unit charge to this gauge multiplet is:
 \eqn\chiralcont{{\cal I}_\chi^R(x;z;m) \equiv (x^{1-R} z^{-1})^{|m|/2} \prod_{j=0}^\infty \frac{1- (-1)^m {z}^{-1} x^{|m| + 2-R + 2 j}}{1-(-1)^m z x^{ |m| + R + 2 j}}.}

For a general gauge theory with gauge group $G$ of rank $r_G$, one introduces fugacities $z_i\;( i=1,\cdots,r_G)$ parameterizing the maximal torus of $G$, with corresponding GNO magnetic fluxes $m_i$ on $\S^2$.  One can similarly introduce fugacities $u_a$ and fluxes $n_a$ for background gauge multiplets coupled to global symmetries.  The $1$-loop determinant in such a configuration is given by taking the product of the contributions \chiralcont\ of the chiral multiplets, along with a contribution from the vector multiplet:
 \eqn\vectorcont{{\cal I}_V(x;z_i;m_i) \equiv \prod_{\alpha \in R} x^{-|\alpha(m)|/2} ( 1 - (-1)^{\alpha(m)} z^{\alpha} x^{|\alpha(m)|}),}
where the product is over the roots of the gauge group.  One can also include Chern-Simons terms for background or dynamical gauge multiplets, whose contribution, for instance, for a level $k$ term for a $U(1)$ gauge multiplet, is $z^{k m}$.  Finally, the partition function is given by integrating over the gauge parameters $z_i$ and summing over the gauge fluxes $m_i$.

We will be interested in the $SO(N_c)$ gauge theory with $N_f$ chiral multiplets $Q_a$ of R-charge $R$ in the vector representation of $SO(N_c)$, and with a Chern-Simons term at level $k$.  We include fugacities and fluxes, $\mu_a$  and $n_a$, $a=1,\cdots,N_f$, for the $U(N_f)$ flavor symmetry, as well as fugacities $\zeta=\pm 1$ for the global symmetry $\Z_2^{\cal M}$, and $\chi=\pm 1$ for the charge conjugation symmetry $\Z_2^\CC$.

Let us write down explicitly the relevant indices.
The index with $\chi=+1$ is given by:
\eqn\threeDindexSONE{
\eqalign{
& {\cal I}^A_{SO(N_c)}(x;\mu_a;n_a;\zeta,\chi=+1) =
\zeta^{k_\zeta \sum_a n_a} \prod_a {\mu_a}^{k_F n_a} \;
\sum_{\{m_1,\dots,m_{r_G}\}} \zeta^{\sum_i m_i} \frac{1}{|W_{\{m_\ell\}}|} \cr
& \oint \prod_{\ell=1}^{r_G}\bigg(\frac{dz_\ell}{2\pi i\,z_\ell}{z_\ell}^{k m_\ell} \bigg)
\,\prod_{a=1}^{N_f}\bigg( ({\cal I}_\chi^R(x;\mu_a;n_a))^\epsilon \prod_{i=1}^{r_G} {\cal I}_\chi^R(x;z_i \mu_a;m_i+n_a){\cal I}_\chi^R(x;{z_i}^{-1} \mu_a;-m_i+n_a) \bigg)\cr
& \prod_{i < j}^{r_G} \bigg( x^{-|m_i-m_j|} ( 1 - (-1)^{m_i-m_j} z_i {z_j^{-1}} x^{|m_i-m_j|}) ( 1 - (-1)^{m_i-m_j} {z_i^{-1} }z_j x^{|m_i-m_j|})  \cr
& \qquad x^{-|m_i+m_j|} ( 1 - (-1)^{m_i+m_j}  z_i z_jx^{|m_i+m_j|}) ( 1 - (-1)^{m_i+m_j}{z_i^{-1} }{z_j^{-1}} x^{|m_i+m_j|}) \bigg) \cr
&\bigg( \prod_{i=1}^{r_G} x^{-|m_i|} (1 - (-1)^{m_i}z_i x^{|m_i|} )(1 -(-1)^{m_i} {z_i}^{-1} x^{|m_i|}) \bigg)^{\epsilon}.
}}
Here $N_c=2r_G+\epsilon$ with $\epsilon=0,1$. The integers $m_i$ run over the Weyl-inequivalent GNO charges, and $|W_{\{m\}}|$ is the order of the residual Weyl group \refs{\KimWB}. In the first term on the first line we have introduced background Chern-Simons terms for the global symmetries. $k_F$ is the level of a background Chern-Simons term for the $U(N_f)$ global symmetry \refs{\ClossetVG,\ClossetVP}\foot{We could also introduce different levels for $U(1)_A$ and $SU(N_f)$.}, and $k_\zeta$ (obeying $k_\zeta \sim k_\zeta + 2$) is a similar term mixing the discrete $\Z_2^{\cal M}$ symmetry with $U(1)_A$\foot{If we describe the background $\Z_2^{\cal M}$ gauge symmetry by two $U(1)$ gauge fields $A_{1,2}$ with an action given by an off-diagonal Chern-Simons term at level two \refs{\MaldacenaSS,\BanksZN}, we can write $k_\zeta$ as the coefficient of an ordinary Chern-Simons term that mixes $A_1$ with the background $U(1)_A$ gauge field.}. We are free to choose the values of these terms, as long as parity anomalies are canceled so that the index is well defined, namely, it has an expansion in fugacities with integer powers.\foot{Note that, in the building blocks defining the index, there appear half-integer powers of the fugacities (see, {\it e.g.}~\chiralcont), and so these factors are not well-defined individually.  However, when we expand the index as a series and include the appropriate background Chern-Simons terms, this expansion has only integer powers of the fugacities, and thus is well-defined.
} This requires
\eqn\demand{
k_\zeta\in \{0,1\}\,,\qquad\quad
k_F\in \Z+\frac12 \, N_c\,,
}
where the second requirement is the standard parity anomaly \refs{\NiemiRQ,\RedlichDV}.

Next we want to compute the index with $\chi=-1$, where we should sum over holonomies of $O(N_c)$ that have determinant $(-1)$. The computation
is different for the cases of odd and even $N_c$~(see~\refs{\HwangQT,\HwangHT} and also \ZwiebelWA).
First, let  us discuss the odd $N_c$ case.
A general $O(2r_G+1)$ holonomy of determinant $\chi$ can be brought to the form
\eqn\diagodd{{\rm diag}\left(z_1,\,z_1^{-1},\,\cdots,z_{r_G},\,z_{r_G}^{-1},\,\chi\right)\,.}
 Thus, the indices with $\chi=-1$ are given by\foot{
In the expression of the index here and in the even $N_c$ case below, we write the last eigenvalue of the holonomy as
$\chi$ in the contribution of the chiral multiplets, in order to keep track of the fractional powers of $\chi$ appearing in the intermediate expressions
in a consistent way.
}
\eqn\threeDtildeindexSONE{
\eqalign{
& {{\cal I}}^A_{SO(N_c)}(x;\mu_a;n_a;\zeta,\chi=-1) =
\left(\zeta^{k_\zeta}\chi^{k_\chi}\right)^{\sum_a n_a} \prod_a {\mu_a}^{k_F n_a}
\;
\sum_{\{m_1,\dots,m_{r_G}\}} \zeta^{\sum_i m_i} \frac{1}{|W_{\{m_\ell\}}|} \cr
& \oint \prod_{\ell=1}^{r_G}\bigg(\frac{dz_\ell}{2\pi i\,z_\ell}{z_\ell}^{k m_\ell} \bigg)
\,\prod_{a=1}^{N_f}\bigg( {\cal I}_\chi^R(x;\chi \mu_a;n_a)\prod_{i=1}^{r_G} {\cal I}_\chi^R(x;z_i \mu_a;m_i+n_a){\cal I}_\chi^R(x;{z_i}^{-1} \mu_a;-m_i+n_a) \bigg)\cr
& \prod_{ i < j}^{r_G} \bigg( x^{-|m_i-m_j|} ( 1 - (-1)^{m_i-m_j} z_i {z_j^{-1}} x^{|m_i-m_j|}) ( 1 - (-1)^{m_i-m_j} {z_i^{-1} }z_j x^{|m_i-m_j|})  \cr
& \qquad x^{-|m_i+m_j|} ( 1 - (-1)^{m_i+m_j}  z_i z_jx^{|m_i+m_j|}) ( 1 - (-1)^{m_i+m_j}{z_i^{-1} }{z_j^{-1}} x^{|m_i+m_j|}) \bigg) \cr
&\bigg( \prod_{i=1}^{r_G} x^{-|m_i|} (1 + (-1)^{m_i}z_i x^{|m_i|} )(1 +(-1)^{m_i} {z_i}^{-1} x^{|m_i|}) \bigg)\,.
}}
We introduced here a background Chern-Simons term with coefficient $k_\chi$ ($k_{\chi} \sim k_{\chi} + 2$) mixing the charge conjugation symmetry $\Z_2^\CC$ with $U(1)_A$, and for the partition function to be well-defined we must have
\eqn\kchi{k_{\chi} \in \{-\frac{1}{2}, \frac{1}{2}\}\,.}
For even $N_c=2\,r_G$, any holonomy of determinant $\chi=-1$ can be brought to the form
\eqn\diageven{{\rm diag}\left(z_1,\,z_1^{-1},\,\cdots,z_{r_G-1},\,z_{r_G-1}^{-1},\,1,\,-1\right)\,.}
Thus, the index is given by
\eqn\threeDindexSONEchi{
\eqalign{
& {{\cal I}}^A_{SO(N_c)}(x;\mu_a;n_a;\zeta,\chi=-1) =
\left(\zeta^{k_\zeta}\chi^{k_\chi}\right)^{\sum_a n_a} \prod_a {\mu_a}^{k_F n_a}\;
\sum_{\{m_1,\dots,m_{r_G-1}\}} \zeta^{\sum_i m_i} \frac{1}{|W_{\{m_\ell\}}|} \cr
& \oint \prod_{\ell=1}^{r_G-1}\bigg(\frac{dz_\ell}{2\pi i\,z_\ell}{z_\ell}^{k m_\ell} \bigg)
\,\prod_{a=1}^{N_f}\bigg( \prod_{i=1}^{r_G-1}
\biggl({\cal I}_\chi^R(x;z_i \mu_a;m_i+n_a){\cal I}_\chi^R(x;{z_i}^{-1} \mu_a;-m_i+n_a) \biggr)\cr
&{\cal I}_\chi^R(x;\mu_a;n_a){\cal I}_\chi^R(x;\chi \mu_a;n_a)\bigg)\,
\prod_{j=1}^{r_G-1}
x^{-2|m_j|} ( 1 -  {z_j^{-2}} x^{2|m_j|})\;( 1 -  {z_j^{2}} x^{2|m_j|})\;
\cr
& \prod_{i < j<r_G}^{r_G-1} \bigg( x^{-|m_i-m_j|} ( 1 - (-1)^{m_i-m_j} z_i {z_j^{-1}} x^{|m_i-m_j|}) ( 1 - (-1)^{m_i-m_j} {z_i^{-1} }z_j x^{|m_i-m_j|})  \cr
& \qquad \qquad x^{-|m_i+m_j|} ( 1 - (-1)^{m_i+m_j}  z_i z_jx^{|m_i+m_j|}) ( 1 - (-1)^{m_i+m_j}{z_i^{-1} }{z_j^{-1}} x^{|m_i+m_j|}) \bigg) \,,
}}
with the same quantization conditions on the background Chern-Simons terms as above.

The dual theory has an $SO(\tilde N_c=N_f+|k|-N_c+2)$ gauge group and Chern-Simons level $(-k)$, with $N_f$ chiral multiplets $q_a$ in the vector representation, $N_f(N_f+1)/2$ singlet mesons $M_{ab}$, and, for $k=0$, also a singlet $Y$.  The index of the dual theory with $\chi=1$, with the charges and the background terms mapped appropriately across the duality, is:
\eqn\threeDindexSONM{
\eqalign{
& {\cal I}^B_{SO(\tilde N_c)}(x;\mu_a;n_a;\zeta,\chi=+1) =
\zeta^{(k_\zeta-k_{\chi}) \sum_a n_a} \prod_a {\mu_a}^{k_F n_a}
\;
\cr
&\qquad\zeta^{-\frac{{\rm sign}(k)}{2} \sum_a n_a} \bigg( \prod_a (\mu_a x^R)^{\frac{1}{2} (k n_a -{\rm sign}(k) \sum_a n_a)}\bigg) x^{\frac{{\rm sign}(k)}{2}(N_f+1-N_c) \sum_a n_a} \cr
&\qquad \bigg(\prod_{a\leq b}^{N_f} {\cal I}_\chi^{2R}(z;\mu_a \mu_b,n_a+n_b) \bigg)\;\;
 \left[{\cal I}_\chi^{N_f+2-N_c-N_f R}(z;\zeta^{-1}\prod_a {\mu_a}^{-1},-\sum_a n_a)\right]^{\delta_{k,0}} \cr
& \sum_{\{m_1,\dots,m_{\tilde r_G}\}} \zeta^{\sum_i m_i}\frac{1}{|W_{\{m_\ell\}}|} \oint \prod_{\ell=1}^{\tilde r_G} \bigg( \frac{dz_\ell}{2\pi i\,z_\ell}{z_\ell}^{-k m_\ell} \bigg) \cr
&\,\prod_{a=1}^{N_f}\bigg( ({\cal I}_\chi^{1-R}(x;{\mu_a}^{-1},-n_a))^{\tilde \epsilon} \prod_{i=1}^{\tilde r_G} {\cal I}_\chi^{1-R}(x;z_i {\mu_a}^{-1};m_i-n_a){\cal I}_\chi^{1-R}(x;{z_i}^{-1} {\mu_a}^{-1};-m_i-n_a) \bigg)\cr
& \prod_{i < j}^{\tilde r_G} \bigg( x^{-|m_i-m_j|} ( 1 - (-1)^{m_i-m_j} z_i {z_j^{-1}} x^{|m_i-m_j|}) ( 1 - (-1)^{m_i-m_j} {z_i^{-1} }z_j x^{|m_i-m_j|})  \cr
& \qquad x^{-|m_i+m_j|} ( 1 - (-1)^{m_i+m_j}  z_i z_j x^{|m_i+m_j|}) ( 1 - (-1)^{m_i+m_j}{z_i^{-1} }{z_j^{-1}} x^{|m_i+m_j|}) \bigg)  \cr
&\bigg( \prod_{i=1}^{\tilde r_G} x^{-|m_i|} (1 - (-1)^{m_i} z_i x^{|m_i|} )(1 -(-1)^{m_i} {z_i}^{-1} x^{|m_i|}) \bigg)^{\tilde \epsilon},
}}
where again $\tilde N_c =N_f+|k|-N_c+2=2{\tilde r}_G+{\tilde \epsilon}$.
Note that the parameter $\zeta$ now also appears in the contribution of the elementary field $Y$, since the $\Z_2^{\tilde {\cal M}}$ symmetry in theory B also acts on this singlet.  The factors on the first and the second line represent the contribution of background Chern-Simons terms. The background Chern-Simons terms on
the second line are the relative ones, which must be included when $k \neq 0$ \BeniniMF\ (here we defined ${\rm sign}(k)=0$ for $k=0$)
for the duality to work.
The expressions for ${\cal I}^B_{SO({\tilde N}_c)}$ with $\chi=-1$ are obtained in an analogous way to our discussion of theory A above.

The dualities discussed in this paper imply the following equality for the indices,
\eqn\equalities{
{\cal I}^A_{SO(N_c)}(x;\mu_a;n_a;\zeta,\,\chi)=
 {\cal I}^B_{SO(N_f+|k|-N_c+2)}(x;\mu_a;n_a;\zeta,\zeta\,\chi)\,.
}
We have checked this equality for various values of the discrete parameters $k$, $n_a$, $\zeta$ and $\chi$, by expanding both sides in a power series in $x$ and comparing the leading coefficients.

We also can write the indices for other orthogonal gauge groups. In the $SO(N_c)$ index computation we introduced a fugacity $\chi=\pm 1$ for the global charge conjugation symmetry $\Z_2^\CC$. Similarly, we can introduce in the computation for an $O(N_c)$ gauge group a discrete theta-like parameter $\chi' = \pm 1$, determining whether we project on even or odd states under $\Z_2^\CC$. The $O(N_c)_+$ result for $\chi'=1$ is half of the sum of the $SO(N_c)$ results with $\chi=1,-1$, and the $O(N_c)_+$ result for $\chi'=-1$ is half of their difference. Thus, allowing for arbitrary $\chi$ and $ \chi'$ one can relate the $SO(N_c)$ and the $O(N_c)_+$ expressions.\foot{This is a discrete version of the situation for $U(N_c)$ gauge groups discussed in \refs{\KapustinSim,\ARSW,\ParkWTA}, where one can go between $SU(N_c)$ and $U(N_c)$ by gauging a $U(1)$ or by ``ungauging'' a topological $U(1)_J$ symmetry.}
In the $O(N_c)_-$ case we need to change the sign of the projection for states charged under $\Z_2^{\cal M}$.
Similarly, the $Spin(N_c)$ and $Pin(N_c)$ indices are given by summing over the sectors with different $\zeta$, and we can define for them an index with $\zeta'=1$ that projects on the $\Z_2^{\cal M}$-even states (which make up the standard $Spin(N_c)$ and $Pin(N_c)$ theories), and an index with $\zeta'=-1$ that projects on the odd states. We then have:
\eqn\spino{\eqalign{
&{\cal I}_{Spin(N_c)}(x;\mu_a;n_a;\zeta',\chi)=\cr
&\qquad\qquad\qquad\frac12\left({\cal I}_{SO(N_c)}(x;\mu_a;n_a;\zeta=+1,\chi)+\zeta'~{\cal I}_{SO(N_c)}(x;\mu_a;n_a;\zeta=-1,\chi)\right)\,,\cr
&{\cal I}_{O(N_c)_+}(x;\mu_a;n_a;\zeta,\chi')=\cr
&\qquad\qquad\qquad\frac12\left({{\cal I}}_{SO(N_c)}(x;\mu_a;n_a;\zeta,
\chi=+1)+ \chi'~{{\cal I}}_{SO(N_c)}(x;\mu_a;n_a;\zeta,\chi=-1)\right)\,,\cr
&{\cal I}_{O(N_c)_-}(x;\mu_a;n_a;\zeta,\chi')=\cr
&\qquad\qquad\qquad\frac12\left({{\cal I}}_{SO(N_c)}(x;\mu_a;n_a;\zeta,
\chi=+1)+ \chi'~{{\cal I}}_{SO(N_c)}(x;\mu_a;n_a;-\zeta,\chi=-1)\right)\,,\cr
&{\cal I}_{Pin(N_c)}(x;\mu_a;n_a;\zeta',{\chi}')=\cr
&\qquad\qquad\qquad\frac12\left({{\cal I}}_{Spin(N_c)}(x;\mu_a;n_a;\zeta',\chi=+1)+  \chi'~{{\cal I}}_{Spin(N_c)}(x;\mu_a;n_a;\zeta',\chi=-1)\right)\,.
}
}
Our tests of the $SO(N_c)$ duality \equalities\ with general $\zeta$ and $\chi$ provide tests also for the dualities between two $O(N_c)_+$ theories, between $Spin(N_c)$ and $O({\tilde N}_c)_-$ theories, and between two $Pin(N_c)$ theories.\foot{Note that the standard indices of these theories map to each other,
but the twisted indices which project onto odd states map only up to a sign, because of the non-trivial
mapping of the $\Z_2$ global symmetries that is embodied in \equalities.}

When the indices are expanded as a series in the various fugacities, the terms in the series represent the contributions of BPS operators with the corresponding charges, as follows from
the definition of the index~\indexDefthree.  Thus we can use the expressions above to attempt to trace how the baryons map between the two dual $SO$ descriptions \equalities.\foot{This statement comes with an
obvious caveat. The index counts operators with signs in such a way that long multiplets cancel out, so we cannot exclude the possibility that there may be multiple states contributing to a given term. } On the electric side the usual baryons are $Q^{N_c}$, with an anti-symmetric product of some choice of $N_c$ flavors. For example, let us define $B$ to be the operator constructed out of the flavors $a=1,\cdots,N_c$.  In the dual side, this maps to a monopole-baryon operator, $\widetilde \beta'$, as described in section \csdual.  Specifically, in this case:
\eqn\barmap{ B = \prod_{a=1}^{N_c} Q_a  \qquad\to\qquad \widetilde \beta'=e^{\tilde \sigma_1/{\hat g}_3^2+i\tilde a_1} \,\left(\prod_{\alpha=3}^{k+2} \tilde \lambda_{1\alpha}\right)\, \left( \prod_{a=N_c+1}^{N_f} q_a\right)\,.}
Here $e^{\tilde\sigma_1/{\hat g}_3^2+i\tilde a_1}$ is the basic $SO({\tilde N_c})$ monopole with GNO charges $(1,0,0,\dots,0)$, $\tilde \lambda_{\alpha \beta}$ are the gluinos, and $q_a$ are the dual quarks, with their color indices contracted anti-symmetrically in the subgroup $SO(N_f-N_c) \subset SO(\tilde N_c)$ that is left unbroken by the monopole and the gluinos.  To test this mapping in the index, note that each chiral multiplet $Q_{a,\alpha}$ ($\alpha=1,\cdots,N_c,\; a=1,\cdots,N_f$) contributes a factor\foot{Here we use the notation $w_{2j-1}=z_j$ and $w_{2j}={z_j}^{-1}$, $j=1,\cdots,r_G$, and $w_{N_c}=1$ for $N_c$ odd, where $z_j,\; j=1,\cdots,r_G$ span the maximal torus of $SO(N_c)$ as in \threeDindexSONE.  Thus the $w_\alpha, \alpha=1,\cdots,N_c$ run over the weights of the vector representation, with $\prod_{\alpha=1}^{N_c} w_\alpha = 1$.} $x^R w_\alpha \mu_a$, so the operator $B$ contributes a term $x^{R N_c} \prod_{a=1}^{N_c} \mu_a$ to the index.  On the dual side,
$q_{a,\beta}$ contributes $x^{1-R} \, w_\beta {\mu_a}^{-1}$, each gluino contributes $-x\,w_\alpha/w_1$, and the monopole background contributes $(-1)^k w_1^k \,x^{N_c-N_f-k+ N_f R} \prod_{a=1}^{N_f} \mu_a$. Putting this together we find that the contribution of $\tilde \beta'$ matches that of $B$ (there is also a factor of $\chi$ in theory A, and a factor of $\zeta \chi$ in theory B).  The baryons of theory B $q^{\tilde N_c}$ are mapped in a similar way to $\beta' = e^{\sigma_1/{\hat g}_3^2+ia_1} \,\left(\prod_{\alpha=3}^{k+2} \lambda_{1\alpha}\right)\,Q^{N_c-k-2}$ in theory A.

\subsec{The partition functions on $\S^3$ }

Let us now comment on the $\S^3$ partition functions.
The partition functions on $\S^3$ for ${\cal N}=1$ SQCD with $O(N_c)_+$ gauge group
were computed and found to agree for the $O(N_c)_+$ dualities discussed in~\refs{\KapustinGH,\HwangQT,\BeniniMF,\HwangHT}.
 In fact, the equality  of the partition functions of the theories with $SO(N_c)$ (and $Spin(N_c)$) gauge groups
discussed in the preceding sections follows directly from the
equality of partition functions for theories with $O(N_c)_+$ gauge groups. The $\S^3$ partition functions
are computed by a matrix integral over the Lie algebra, which is the same in all these cases,
and thus the partition functions differ only by  overall factors of $2$ due to the different volumes of the gauge groups.
Hence, the results of~\refs{\KapustinGH,\HwangQT,\BeniniMF,\HwangHT} straightforwardly imply that the $SO(N_c)$ dual pairs
discussed in this paper have the same $\S^3$ partition functions.

In certain cases, {\it e.g.} the dualities discussed in~\ARSW,
the equality of the  partition functions on $\S^3$ of the $3d$ theories follows in a simple way
from the equality of the $4d$ partition functions on $\S^3\times \S^1$ (the supersymmetric index)
of the $4d$  theories from which these $3d$ theories descend.  However, this is more subtle
in the case of dualities with orthogonal groups, as we will now explain.

First, let us briefly outline how the $3d$ partition functions are obtained from the $4d$ indices:
for more details see~\refs{\DolanRP,\GaddeIA,\ImamuraUW,\NiarchosAH,\ARSW}. The partition function of a $4d$ theory on $\S^3\times \S^1$ can be thought of explicitly as an $\S^3$ partition function of the dimensionally reduced theory
with all the KK modes included.\foot{
For the free chiral field, the representation of the $4d$ index as a product of $3d$ partition functions
of KK modes is the physical content of~\slthreeZ, as explained in appendix B of \ARSW.
} The (inverse) radius of the $\S^1$  appears in the $\S^3$
partition function as a real mass for the $U(1)$ symmetry associated
with the rotation around the circle. Taking the small radius limit  corresponds to
taking this real mass to be large, and thus decoupling the massive KK modes.
The fugacities for the $4d$ global symmetries become real mass parameters in $3d$.
Some of the classical symmetries of the  $4d$ gauge theories are anomalous,
but the $3d$ theories obtained by  dimensional reduction of  the matter content of the $4d$ ones
do have these symmetries at the full quantum level. The $4d$ index cannot be refined
with fugacities for the  anomalous symmetries, and thus the $3d$ partition functions obtained
by this reduction procedure are not refined with the corresponding real mass parameters.
This is an indication that the $3d$ theory obtained by the reduction has a superpotential
breaking the symmetry that is anomalous in $4d$ \ARSW.

The above discussion presumes
that the dimensional reduction produces a well-defined and finite $\S^3$ partition function.
This presumption is true for  the cases discussed in~\ARSW, but it is not true for
 the $SO(N_c)$ theories discussed in this paper: the reduction of the $4d$ index\foot{
The equality of $4d$ indices of dualities with $so(N_c)$ Lie algebras were checked
 in~\DolanQI. See also~\SpiridonovHF\ for a related discussion.
}
 for $SO(N_c)$
SQCD produces a divergent $3d$ partition function. The divergence can be explained physically
by the fact that not all of the Coulomb branch is lifted when putting the theory on the circle,
as discussed in the previous sections. In particular, in the $4d$ theory on $\S^1$, the operators $Y$ or $Y_{Spin}$, parameterizing the Coulomb branch, have no continuous global symmetry charges and no R-charge,
and the presence of such a field leads the $3d$ partition function to diverge.\foot{
One can see this problem directly in $3d$. The partition functions on $\S^3$ and $\S^2\times \S^1$, of $SO(N_c)$ SQCD with general real mass parameters and R-charges, are  finite and well defined.
However, on the subspace corresponding to the $4d$ R-symmetry, and where we turn on
only real masses for non-anomalous symmetries in $4d$, both partition functions diverge.
}  We have seen
in the previous sections that due to the intricate moduli space on the circle,
the $3d$ $SO(N_c)$ dualities are obtained by focusing on certain regions of the Coulomb branch.
It is possible that this more intricate procedure can also be mimicked at the
level of the index,\foot{A possible way to obtain a finite $3d$ partition function
is to take a ``double scaling'' limit,
of a small radius of the circle together with taking some of the real masses to be large.}
and we leave this question
to future investigations.

\bigskip

\noindent {\bf Acknowledgments:}

We would like to thank R.~Dijkgraaf, D.~Freed, K.~Intriligator, G.~Moore, J.~Park, I.~Shamir, Y.~Tachikawa and E.~Witten for useful discussions. OA is the Samuel Sebba Professorial Chair of Pure and Applied Physics, and he is supported in part by a grant from the Rosa and Emilio Segre Research Award, by an Israel Science Foundation center for excellence grant, by the German-Israeli Foundation (GIF) for Scientific Research and Development, by the Minerva foundation with funding from the Federal German Ministry for Education and Research, and by the I-CORE program of the Planning and Budgeting Committee and the Israel Science Foundation (grant number 1937/12). OA gratefully acknowledges support from an IBM Einstein Fellowship at the Institute for Advanced Study. SSR gratefully acknowledges support from the Martin~A.~Chooljian and Helen Chooljian membership at the Institute for Advanced Study. The research of SSR was also partially supported by NSF grant number PHY-0969448. The work of NS was supported in part by DOE grant DE-SC0009988 and by the United States-Israel Binational Science Foundation (BSF) under grant number~2010/629. The research of BW was supported in part by DOE Grant DE-SC0009988.

\def\listrefs{
\immediate\closeout\rfile\writestoppt
\bigskip\baselineskip=\footskip{\noindent {\bf References}\hfill}\medskip{\parindent=20pt%
\frenchspacing\escapechar=` \input \jobname.refs\vfill\eject}\nonfrenchspacing}

\

\

\

\

\

\listrefs
\end